\documentclass[aps,prd,twocolumn,showpacs,nofootinbib,showkeys,superscriptaddress,preprintnumbers,floatfix]{revtex4-2}

\usepackage{amsmath,amssymb,graphicx,yfonts,tensor}
\usepackage{adjustbox}
\usepackage{hyperref}
\usepackage{color}
\graphicspath{{figures/}}
\newcommand{\tensd}[2]{#1\indices{_{#2}} }
\newcommand{\tensu}[2]{#1\indices{^{#2}} }
\newcommand{\tensud}[3]{#1\indices{^{#2}_{#3}}}
	
\newcommand{\pr}{\partial}
\newcommand{\be}{\begin{equation}}
\newcommand{\ee}{\end{equation}}
\newcommand{\bec}{\begin{equation*}}
\newcommand{\eec}{\end{equation*}}
\newcommand{\eol}{ \\
	&\phantom{=} 
}
\newcommand{\nabmin}{\nabla^{(\eta)}}
\newcommand{\hp}{h_{+}}
\newcommand{\hx}{h_{\times}}
\newcommand{\bea}{\begin{eqnarray}}
\newcommand{\eea}{\end{eqnarray}}
\begin{document}
\title{Detecting~Planetary-mass~Primordial~Black~Holes with~Resonant~Electromagnetic~Gravitational~Wave~Detectors }

\author{Nicolas Herman} \email{nicolas.herman@unamur.be}
\affiliation{ Department of Mathematics and Namur Institute for Complex Systems (naXys), University of Namur, Rue Grafé 2, B-5000, Namur, Belgium}

\author{Andr\'e F\H{u}zfa} \email{andre.fuzfa@unamur.be}
\affiliation{ Department of Mathematics and Namur Institute for Complex Systems (naXys), University of Namur, Rue Grafé 2, B-5000, Namur, Belgium}
\affiliation{Cosmology, Universe and Relativity at Louvain, Institute of Mathematics and Physics,
Louvain University, 2 Chemin du Cyclotron, 1348 Louvain-la-Neuve, Belgium}

\author{L\'eonard Lehoucq}
\email{leonard.lehoucq@ens-paris-saclay.fr}
\affiliation{ Department of Mathematics and Namur Institute for Complex Systems (naXys), University of Namur, Rue Grafé 2, B-5000, Namur, Belgium}
\affiliation{Department of theoretical physics at the ENS Paris-Saclay, University of Paris-Saclay, avenue des Sciences, 91190, Gif-sur-Yvette, France}

\author{Sebastien Clesse}  \email{sebastien.clesse@ulb.ac.be}
\affiliation{Service de Physique Th\'eorique, Universit\'e Libre de Bruxelles (ULB), Boulevard du Triomphe, CP225, B-1050 Brussels, Belgium}
\affiliation{Cosmology, Universe and Relativity at Louvain, Institute of Mathematics and Physics,
Louvain University, 2 Chemin du Cyclotron, 1348 Louvain-la-Neuve, Belgium}



\date{\today}

\begin{abstract}
\noindent   The possibility to detect gravitational waves (GW) from planetary-mass primordial black hole (PBH) binaries with electromagnetic (EM) detectors of high-frequency GWs is investigated.  We consider two patented experimental designs, based on the inverse Gertsenshtein effect, in which incoming GWs passing through a static magnetic field 
induce EM excitations inside either a TM cavity or a TEM waveguide.  The 
frequency response of the detectors is computed for 
post-newtonian GW waveforms.   We find that such EM detectors based on  current technology may achieve a strain sensitivity down to $h \sim 10^{-30}$, which generates an EM induced power 
of $10^{-10}$ W.   This allows the detection of PBH binary mergers of mass around $10^{-5} M_\odot$ if they constitute more than $0.01$ percent of the dark matter, as suggested by recent microlensing observations.  We envision that 
this class of detectors could also be used to detect cosmological GW backgrounds and probe sources in the early Universe at energies up to the GUT scale.
\end{abstract}

\pacs{}
\maketitle

\section{Introduction}

Gravitational Waves (GW), first introduced by Einstein in 1916 \cite{einstein1,einstein2} as a linear regime of the field equations of General Relativity, have been directly detected for the first time in 2015 by the LIGO/Virgo collaboration~\cite{LIGO}.   These ground-based detectors use laser interferometry techniques, but there also exists other detection strategies.   Einstein's Equivalence Principle implies that all types of energies produce and experience gravity in the same way.   The energy of electromagnetic (EM) radiation must therefore source gravity just like compact objects do and, the other way around, gravity (e.g. gravitational waves) can manifest itself in the physical characteristics of electromagnetic radiation.   This is the basic principle behind the EM detection (or emission) of GWs.  

The use of EM fields to both generate and detect GWs has been actually considered for decades, for instance based on the (inverse) Gertsenshtein effect~\cite{gertsenshtein} relying on the coupling between GWs and EM waves in the presence of a strong static magnetic field.  However, the weakness of this coupling makes any GW detection extremely challenging.  Therefore experimental efforts have mostly concentrated on ground-based detectors based on laser interferometry, probing a rather low frequency range (typically $1-10\, {\rm kHz}$).  But this technique is not suited for the detection of high-frequency GWs (HFGWs), at the opposite of EM-based detectors.    The Section~\ref{sec:hist} presents a short historic review of HFGW detectors that have been build or proposed.  The interested reader will find more details about detection strategies and potential high-frequency sources in a recent review~\cite{Aggarwal:2020olq}.

In this paper, we propose two novel patented~\cite{patent} experimental designs of resonant high-frequency EM detectors, based on the inverse Gerstenshtein effect, operating at MHz or GHz frequencies and feasible with current technology.   We compute numerically the EM signal produced by passing HFGWs and consider planetary-mass primordial black holes (PBHs) as their potential sources.  The detectors are constituted by either a waveguide or a cavity immersed into a transverse static magnetic field.  This outer magnetic field deserves two purposes in our design. First, a proper (transverse) orientation of this field is mandatory to convert HFGWs into EM waves through the inverse Gerstenshtein mechanism. Second, the external magnetic field boosts the output signal through a resonance mechanism on specific radiation modes that are excited by the passing HFGWs.

There exists a broad range of hypothetical astrophysical or cosmological sources of HFGWs with frequencies above kHz (see~\cite{Aggarwal:2020olq} and references therein), such as exotic compact objects, PBHs, inflation, reheating, oscillons, cosmic strings and other topological defects, large curvature fluctuations and phase transitions in the early Universe.   Some sources produce transient signals, like the merging of compact objects, while cosmological sources typically induce a permanent stochastic GW background.  In particular, HFGW detectors operating  at MHz and GHz frequencies might probe new Physics up to the scale of the Grand Unified Theories.

In this paper, we focus on the particular case of transient signals from the merging of light (planetary-mass scale) PBHs.   Their existence is still hypothetical but motivated by several recent observations, such as microlensing events towards the galactic bulge, recently detected by OGLE~\cite{Niikura:2019kqi,Mroz2017} and suggesting that such PBHs may constitute between 1\% and 10\% of the Dark Matter (DM).   LIGO/Virgo observations provide additional motivations for the existence of PBHs in the stellar range~\cite{Bird:2016dcv,Clesse:2016vqa,Sasaki:2016jop} and a unified scenario with a wide mass distribution imprinted by the known thermal history of the Universe has been presented in~\cite{Carr:2019kxo}.   PBHs may originate from the gravitational collapse of primordial inhomogeneities in the early Universe.  Above a mass $m_{\rm PBH} \approx10^{11}$ kg, their evaporation time through the Hawking-Bekenstein mechanism is much larger than the age of the Universe.   Beyond these motivations, the GW waveform from PBH mergers is well-known, which allows us to simulate the exact detector response. 
Our work therefore aims at paving the road towards an experimental realization of a HFGW detector with a good enough strain sensitivity, of order $h\sim 10^{-30}$, to detect various GW sources.  

The paper is organized as follows:   After reviewing previous proposals of EM-based HFGW detectors in Section~\ref{sec:hist}, we introduce in Section~\ref{sec:EM-G} the theory behind the Einstein-Maxwell system and the Gertsenshtein effects.  Section~\ref{sec:det} applies the theory to the case of two detector designs based on a resonant waveguide and a cavity, into an external static magnetic field.  Section~\ref{sec:PBH} is devoted to the computation of the expected rate and signal from planetary-mass PBH mergers.   In Section~\ref{sec:det-PBH} we gather these calculations to compute the expected signal in our detectors, in terms of frequency response and induced EM power, for different realistic designs and different PBH masses.  Finally, we compute forecasted limits on the possible PBH abundance, for two possible binary formation channels (primordial binaries and tidal capture in clusters).   We discuss our results and present some perspectives in the conclusion (Section~\ref{sec:conc}).

\section{Brief History of Electromagnetic and HFGW detectors} \label{sec:hist}
The use of EM fields to both generate and detect GWs has been actually considered for decades. While Weber~\cite{weber} envisioned the importance of both the generation and detection of GWs as early as 1960, Gertsenshtein~\cite{gertsenshtein} discovered in 1962 a resonance mechanism allowing to produce GWs from EM waves in the presence of a strong static magnetic field.  Later in~\cite{boccaletti,delogi}, this mechanism was studied in greater details using Einstein's equations, scattering theory and Feynman perturbation techniques.  Gertsenshtein's mechanism was applied to astrophysics by Zeldovich~\cite{zeldovich}. Grishchuk and Sazhin then introduced in~\cite{grishchuk1,grishchuk2,grishchuk2bis} purely electromagnetic generators of GWs, using transverse magnetic/electric (TM/TE) resonant cavities.   They envisioned GW emission-reception laboratory experiments and  concluded that they might be experimentally feasible~\cite{grishchuk2,grishchuk2bis}, which would open the road to futuristic technologies based on GW physics~\cite{grishchuk3}.  Resonant cavities and EM waveguides were then considered as possible detectors of gravitational radiation, either emitted by natural or artificial sources~\cite{braginskii1,braginskii2,grishchuk2,grishchuk2bis,grishchuk3,pegoraro1,pegoraro2,caves,gerlach}.

EM detectors of GWs would allow exploring higher frequency range than with laser interferometry, typically from kHz to 100 GHz when using radio-frequencies or from 100GHz to THz when using microwaves. 
Investigations of electromagnetic GW detectors began in the 1970's with the works~\cite{braginskii1,braginskii2,grishchuk2,grishchuk2bis,boccaletti,pegoraro1,pegoraro2,caves,cruise}. Those detectors either make use of the conversion of GWs to photons \cite{boccaletti}, the excitation or modification of resonant modes of EM cavities and waveguides~\cite{braginskii1,braginskii2,grishchuk2,grishchuk2bis,pegoraro1,pegoraro2,caves,ballantini}, the change of polarization plane of an EM wave due to the passing GWs~\cite{cruise}, or induced birefringence of the interior of the cavity~\cite{gerlach}. More recently, Ejlli, Cruise et al. in~\cite{Ejlli2019} used available data from experiments designed for the detection of weakly interacting slim particles to set limits on the stochastic GW background through the graviton to photon conversion in the ultra-high frequency band (above 1 THz).  There has been also a method proposed by Bentley et al.~\cite{bentley} to reduce the noise of interferometric GW detectors at high-frequency (kHz).
Other types of HFGW detectors have been recently proposed~\cite{gwdetect1,Aggarwal:2020umq,Ito2020,Aggarwal:2020umq,Goryachev2014}, using optically trapped dielectric microspheres in a cavity, resonance between a graviton and a magnon that is based on the Dirac equation in a curved spacetime or high frequency phonon trapping acoustic cavities.

\section{Einstein-Maxwell system and the Gertsenshtein effects} \label{sec:EM-G}
The Einstein-Maxwell system models the interplay of gravitation and electromagnetism  
by the coupling of their respective field equations (in S.I. units\footnote{The relevant fundamental constants of the Einstein-Maxwell system are $G$ as Newton's constant, $c$ as the speed of light and $\mu_0$ as the (vacuum) magnetic permeability.})
\bea
R_{\mu\nu}&=&\frac{8\pi G}{c^4} T^{\left(\rm em\right)}_{\mu\nu} \; \label{einstein}\\
\nabla_\mu F^{\mu\nu}&=&-\mu_0 J^\nu\; \label{maxwell}
\eea
where 
\be
T^{\left(\rm em\right)}_{\mu\nu}=\frac{1}{\mu_0}\left(g^{\alpha\beta}F_{\mu\alpha}F_{\nu\beta}-\frac{1}{4}g_{\mu\nu}F_{\alpha\beta}F^{\alpha\beta}\right)\label{tmunu_EM}
\ee
 is the Maxwell stress-energy tensor, $F_{\mu\nu}=\partial_\mu A_\nu-\partial_\nu A_\mu$
being the Faraday tensor of the electromagnetic field, $g_{\mu\nu}$ and $A_\mu$ are the metric and the four-vector potential, $\nabla_\mu$ is the covariant derivative with respect to $g_{\mu\nu}$, $R_{\mu\nu}$ is the Ricci tensor and $J^\nu$ the four-current density. In general relativity, the interplay between gravitation and electromagnetism is twofold: first, spacetime is curved by the energy of the electromagnetic field as ruled by Eq.~(\ref{einstein}) while, at the same time, the propagation of the electromagnetic field $F_{\mu\nu}$ is governed by the covariant Maxwell equations in curved spacetime, Eq.~(\ref{maxwell}).
For a small electromagnetic compactness $G E_{\rm EM}/(c^4 L)\ll 1$, where $E_{\rm EM}$ is the EM energy stored in some physical system of length $L$, the gravitational sector of the system can be safely treated in the weak-field limit~\cite{fuzfa}. 

First, one can consider the gravitational perturbations arising from EM sources in the Einstein field equations. Considering EM configurations consisting in 
a superposition of a static field $F^{(s)}_{\mu\nu}$ and a varying one $F^{(v)}_{\mu\nu}$, the quadratic terms in Eq.~(\ref{tmunu_EM}) yield to three general classes of electromagnetically-induced gravitational perturbations around a  Minkowski background
\be
g_{\mu\nu}=\eta_{\mu\nu}+c_{\mu\nu}+w_{\mu\nu}+h_{\mu\nu}\label{decomp}
\ee
where $\eta_{\mu\nu}=\rm diag(-1,+1,+1,+1)$ is the Minkowski metric and $c_{\mu\nu},w_{\mu\nu},h_{\mu\nu}\ll 1$ represent the metric perturbations. 
In the above equation, (i) $c_{\mu\nu}$ represents the static gravitational field generated by some external static magnetic or electric field (from a coil or a capacitor) and arises from the quadratic term in $F^{(s)}_{\mu\nu}$ in~(\ref{tmunu_EM}), while (ii) $w_{\mu\nu}$ is a gravitational wave generated by the varying EM [for which the source is the quadratic term in $F^{(v)}_{\mu\nu}$ in Eq.~(\ref{tmunu_EM})] and (iii) $h_{\mu\nu}$ is another gravitational wave generated by the coupling between the external static field  $F^{(s)}_{\mu\nu}$ and some EM wave $F^{(v)}_{\mu\nu}$ [crossed terms in Eq.~(\ref{tmunu_EM})].  The case (i) has been studied in~\cite{fuzfa} but does not give rise to a GW since the outer EM field is static. Case (ii) has been studied for light propagating pulses in~\cite{tolman,ratzel} and in~\cite{grishchuk1,grishchuk2,grishchuk2bis,tang} for EM waves in resonant hollow cavities\footnote{In~\cite{grishchuk1}, a special case of hollow spherical cavity in an outer radial magnetic field is briefly considered, giving rise to an admixture of terms $w_{\mu\nu}$ (case (ii)) and $h_{\mu\nu}$ (resonance - case (iii)) which were not identified as such nor exploited by the authors.}. Case (iii) actually corresponds to what is called the direct Gertsenshtein effect~\cite{gertsenshtein,zeldovich}.  
This Gertsenshtein effect is a wave resonance mechanism in which light passing through a region of uniform magnetic field, perpendicular to the direction of light propagation, produces GWs. A monochromatic EM wave leads to an outgoing GW of same frequency. Electromagnetic generation of GWs is a very faint process, due to the extreme weakness of gravitational coupling.   Indeed, the metric perturbations $h_{\mu\nu}$ produced through the Gerstenshtein mechanism have an amplitude of order 
\be
h_{\mu \nu} \sim \frac{4 G B_0 E_0 L^2}{c^5\mu_0}~,
\ee
 where $L$ is the size of the region in which the magnetic field and the EM wave interact, and where $B_0$ and $E_0$ are the amplitudes of the static magnetic and varying electric fields respectively. To give an idea, in order to generate a strain $h\approx 10^{-21}$ with $B_0\approx 10$ T and $E_0\approx 1 \, {\rm MV/m}$, one needs 
 a truly astronomical size of the interacting region, $L\approx 10^6\; {\rm km}$. Therefore, while the direct Gertsenshtein effect can be used to build electromagnetic GW generators, its practical application constitutes an extreme experimental challenge.
 
Second, ripples in spacetime can also interact with a static magnetic field to produce an outgoing EM wave. This inverse Gertsenshtein effect is described by the Maxwell equations, Eq.~(\ref{maxwell}), on a perturbed background. An obvious application of this effect is the detection of GWs passing into a transverse static magnetic field, yielding to their conversion into EM waves. One possible way to derive the equations governing the inverse Gertsenshtein effect is to develop the covariant derivative in Eq.~(\ref{maxwell}) at first order in metric perturbations, and to treat these ones as an effective current density.  We follow here a different approach, based on a covariant generalization of the EM wave equations\footnote{These are obtained from the two groups of covariant Maxwell equations
$ \nabla_\mu F^{\mu \nu}= 0$ and
$\nabla_\kappa F_{\mu \nu} + \nabla_\mu F_{\nu \kappa} + \nabla_\nu F_{\kappa \mu}=0$
which can be combined to retrieve Eq.~(\ref{wavenormal}).}
 \be
  \label{wavenormal}
 g^{\alpha\beta}\nabla_\alpha\nabla_\beta \tensd{F}{\mu \nu} + \tensd{R}{\mu \nu \alpha \beta} \tensu{F}{\alpha \beta} + \tensud{R}{\alpha}{\mu} \tensd{F}{\nu \alpha} +
 \tensud{R}{\alpha}{\nu} \tensd{F}{ \alpha \mu} = 0,
 \ee
where  $\tensd{R}{\mu \nu \alpha \beta}$ is the Riemann tensor. This set of equations describes the propagation of EM waves on a curved spacetime.
In the case of a flat Minkowski spacetime, the wave equation Eq.~(\ref{wavenormal}) reduces to the classical wave equation $ g^{\alpha\beta}\nabla_\alpha\nabla_\beta  \tensd{F}{\mu \nu}=0$.

Let us now consider a small perturbation $h_{\mu \nu}$ propagating on a Minkowski background, such that the metric is given by $g_{\mu \nu}=\eta_{\mu \nu} + h_{\mu \nu}$ at first order ($h_{\mu \nu} \ll 1$). 
If this perturbation satisfies the Lorenz gauge condition, $\partial_{\mu} h^{\mu \alpha} =0$, one gets from Eq~(\ref{wavenormal}) the linearized wave equation for the Faraday tensor (see also~\cite{grishchuk2bis}),
\begin{widetext}
\begin{align}
\begin{split}
g^{\alpha\beta}\nabla_\alpha\nabla_\beta  \tensud{F}{(1)}{\mu \nu} &= \tensu{h}{\alpha \kappa} \nabmin_\alpha \nabmin_\kappa \tensud{F}{(0)}{\mu \nu} - \pr_\rho\left(\pr_\mu \tensd{h}{\alpha \nu}-\pr_\nu \tensd{h}{\alpha \mu}\right) \tensu{F}{(0)  \, \rho \alpha} \eol -\left(\pr^\gamma \tensu{h}{\alpha \beta}  + \pr^\alpha \tensu{h}{ \beta \gamma} - \pr^\beta \tensu{h}{\gamma \alpha} \right) \left(\tensd{\eta}{\alpha \mu} \nabmin_\gamma \tensud{F}{(0)}{\nu \beta}-\tensd{\eta}{\alpha \nu} \nabmin_\gamma \tensud{F}{(0)}{\mu \beta}\right)=S_{\mu \nu}.
\label{lineq}
\end{split}
\end{align}
\end{widetext}

In the above equation, we have assumed that the total EM field $F_{\mu \nu}$ is the superposition of some background EM field $\tensud{F}{(0)}{\mu \nu}$ with which the metric perturbation $h_{\mu\nu}$ interacts to produce an EM perturbation $\tensud{F}{(1)}{\mu \nu}$.  In the following, we focus on the magnetic conversion of GWs into photons, i.e. the interaction between a passing GW and an external static magnetic field ($\tensud{F}{(0)}{\mu \nu}$ is purely magnetic) which results in an EM wave emission.  Eq.~(\ref{lineq}) governs the inverse Gertsenshtein effect. 

Under the assumption of a uniform static magnetic field, the first and third source terms in Eq.  (\ref{lineq}), both including $\nabmin_\gamma \tensud{F}{(0)}{\alpha \beta}$, identically vanish, leaving as the only source term the one with second derivatives of the metric perturbations,
\be
S_{\mu\nu}= -\pr_\alpha \left(\pr_\mu \tensd{h}{\beta \nu}-\pr_\nu \tensd{h}{\beta \mu}\right) \tensu{F}{(0)  \, \alpha \beta}\cdot\label{Smunu}
\ee
We can now apply this theory and conceive a specific experiment for the detection of HFGWs, produced e.g. by inspiralling PBHs.

\section{Resonant Electromagnetic detectors of HFGWs} \label{sec:det}

In this section, we describe two detector designs that are based on the patents~\cite{patent}.
The detection principle is based on the inverse Gertsenshtein effect, thus a passing GW interacts with an intense
static magnetic field.  If the direction of the incoming GW is not colinear with the magnetic field, faint transverse EM waves are generated and these can be further amplified by EM resonators. 
The experimental set-up consists of either a waveguide or a cavity whose axis of symmetry is orthogonal to the magnetic field.
 Such a set-up is similar to haloscopes that are used for the search of axions, like the ADMX experiment~\cite{ADMX,ADMX2}, except for the orientation of outer magnetic field.  As shown below, it is mandatory that it is orthogonal to the axis of the cavity/waveguide in order to detect GWs. 

A theorem by Choquet-Bruhat~\cite{choquet} establishes that both direct and inverse Gertsenshtein effects require the condition of orthogonality between the external EM field
and the direction of GW propagation. This theorem starts from the hypothesis that incoming or generated GWs have a Wentzel-Kramers-Brillouin (WKB) form and are of high-frequencies. The equation of propagation in~\cite{choquet} is then obtained after a development in frequency. We propose below a variant of Choquet-Bruhat's demonstration with a development in amplitude instead of frequency.  We first assume that the incoming GW is a plane wave, 
$h_{\mu \nu} = a_{\mu \nu}\,\,e^{i\omega\Phi},$ 
with a general varying phase, $\Phi=\Phi(x^\alpha)$. The goal is to show that the constant EM field must be orthogonal to the direction of propagation of the incoming plane wave, in order to produce an EM wave.  In other words, no EM wave can be generated from the interaction between a constant EM field and a GW, unless the first one is orthogonal to the direction of propagation of the second one.  In order to show this, we demonstrate that
a vanishing source term $S_{\mu \nu} = 0$ in Eqs.~(\ref{lineq}) is equivalent to the condition $ \Phi_\alpha F^{\alpha \mu\, (0)} = q\,\Phi^\mu$ with $q$ a real constant and $\Phi_\alpha = \partial_\alpha \ \Phi $. 
Since we assume that $E^{(0)}$ and $B^{(0)}$ are constants in our problem, the source term is given by Eq.(\ref{Smunu}).
Therefore, the non-zero part of $S_{\mu \nu}$ is the exterior derivative of an effective 4-current density: $J^\mathrm{eff}_{\mu}=\partial_\alpha\,h_{\beta \mu} \, F^{\alpha \beta\,(0)}$. We can thus rewrite our source term as
\begin{subequations}
	\nonumber
	\begin{align}
	S_{\mu \nu} &= \partial_\nu\,J^\mathrm{eff}_{\mu} - \partial_\mu\,J^\mathrm{eff}_{\nu}\\
	&= (\Phi_\nu\,J^\mathrm{eff}_{\mu} - \Phi_\mu\,J^\mathrm{eff}_{\nu})' \quad \mathrm{with} \quad (\cdot)'\equiv \partial(\cdot)/\partial\Phi \, .
	\end{align}
\end{subequations}
The last line is obtained as a result of the plane wave approximation (the amplitude $a_{\mu\nu}$ above is constant), which implies that the partial derivative $ \partial_\nu $ is equivalent to $\Phi_\nu\,\frac{\partial}{\partial \Phi}$. Because a GW verifies the eikonal $\Phi_\mu \Phi^\mu = 0$ (if not, it is inconsistent and vanishes with a change of coordinate \cite{choquet}), multiplying the above expression by $\Phi^\mu$ gives that $S_{\mu \nu} = 0 \Leftrightarrow \Phi^\mu J^\mathrm{eff}_{\mu} = 0$, and thus $\Phi_\nu$ is orthogonal to $J^\mathrm{eff}_{\mu}$. Since $\Phi_\mu$ does not vanish, this yields to the equivalence $S_{\mu \nu} = 0 \Leftrightarrow J^\mathrm{eff}_{\mu} = 0$.
Let us now show the central result
\be
S_{\mu \nu} = 0 \Leftrightarrow J^\mathrm{eff}_{\mu} = 0 \Leftrightarrow \Phi_\alpha F^{\alpha \mu\, (0)} = q\,\Phi^\mu\label{thm}~.
\ee
For plane waves $h_{\mu \nu} = a_{\mu \nu}\,\,e^{i\omega\Phi(x^\alpha)}$, we have that $\partial_{\alpha} h_{\lambda \nu}= \Phi_{\alpha} h'_{\lambda \nu}$. Therefore, the effective four-current can be simplified
to
\begin{subequations}
	\nonumber
	\begin{align}
		J^\mathrm{eff}_{\ \nu} = F^{\alpha \lambda\, (0)}\, \Phi_{\alpha} \, h'_{\lambda \nu}
	\end{align}
\end{subequations}
If $\Phi_\alpha F^{\alpha \mu \, (0)} = q\,\Phi^\mu$ then $J^\mathrm{eff}_{\ \nu} = q\,\Phi^{\lambda}\,{h'}_{\lambda \nu}.$ In the meantime, the Lorenz gauge  condition, $\partial^\mu \,h_{\mu \nu} = 0$, is equivalent to $\Phi^{\mu}\,{h'}_{\mu \nu} = 0$  in the plane wave approximation. So we can conclude that the effective 4-current density $J^\mathrm{eff}_{\ \nu}$ vanishes . Now, let us now prove the implication in the reverse way.

Let us assume $J^\mathrm{eff}_{\ \nu} = 0$ and move to radiative coordinates, that is to say that $\Phi = x^0$ so we have directly $\Phi_0 = 1$ and $\Phi_i = 0$. The GW obey to the eikonal, thus $\eta^{00} = 0$ and $\eta^{0i} = \Phi^i$. In such a comobile coordinate system, the significant component of the wave are the $h_{ij}$. The source term becomes
\begin{subequations}
	\nonumber
	\begin{align}
	J^\mathrm{eff}_{\ \nu} &= \Phi_0\,F^{0\lambda \, (0)}\,h'_{\lambda \nu} + \Phi_i\,F^{i\lambda \, (0)}\,h'_{\lambda \nu}\\
	&= F^{0\lambda \, (0)}\,h'_{\lambda \nu}\\
	&= F^{0j \, (0)}\,h'_{ij}
	\end{align}
\end{subequations}
So the fact  that $J^\mathrm{eff}_{\ \nu} = 0$ implies that $F^{0j \, (0)}\,h'_{ij} = 0$.
At any point of space time, we can choose a spatial coordinate system such that $F^{0j\,(0)} = A \, \delta^j_1$. Then we have $F^{0j \, (0)}\,h'_{ij} = 0,$ that leads to $A\,{h'}_{i1} = 0 $ so $A = 0$ since the GW is non-zero. So $F^{0j \, (0)} = 0 =q\,\Phi^j $, because $\Phi^j = \eta^{0j} = 0$. Knowing that $\Phi_0 = 1$ and $\Phi_i = 0$, we can show that $F^{0j \, (0)}=\Phi_\alpha \, F^{\alpha j \, (0)}$. We can also consider that $\Phi^0 = 0 $ and $F^{\alpha 0} = - F^{0 \alpha}$ to conclude that $\Phi_\alpha F^{\alpha \mu\, (0)} = q\,\Phi^\mu$. This completes our demonstration.
Let us now particularise the final result $\Phi_\alpha F^{\alpha \mu\, (0)} = q\,\Phi^\mu$ in a illustrating case.
 In cartesian coordinates, we can write down the null vector $\Phi_\alpha = (k,0,0,k)$ with $k$ the wave vector of the incident GW which is therefore propagating along the $z$-direction. The above-mentioned condition Eq.(\ref{thm}) now leads now to two constraints on the EM field : $E^x + B^y = 0$ and $E^y - B^x = 0$. If one considers the case when there is no electric field, then this condition implies that the components of the outer magnetic field that are transverse to the direction of GW propagation vanish: $B^x=B^y=0\cdot$ In other words, any longitudinal magnetic field $B^z$ does not produce any EM wave by inverse Gertsenshtein effect (since $S_{\mu\nu}= 0$). To produce GW by this mechanism, one needs $S_{\mu\nu}\ne 0$ or, equivalently, a non vanishing magnetic field in the direction transverse to the GW propagation ($B_\perp\ne 0$).
 
 This is the reason why experiments like ADMX \cite{ADMX} do not have the right configuration to detect GW. Indeed, they use a longitudinal outer magnetic field which can therefore only interact with GW propagating transversely to it. This interaction can only produce EM waves that are in the same direction as the constant magnetic field but this is forbidden in the
  TM cavity they are using (since only transverse excitation modes are allowed, not longitudinal ones). To turn a haloscope into a HFGW detector, then one simply needs to rotate the outer magnetic field by a quarter of turn.

\begin{figure}[ht!]
\centering
\includegraphics[clip,scale=0.15]{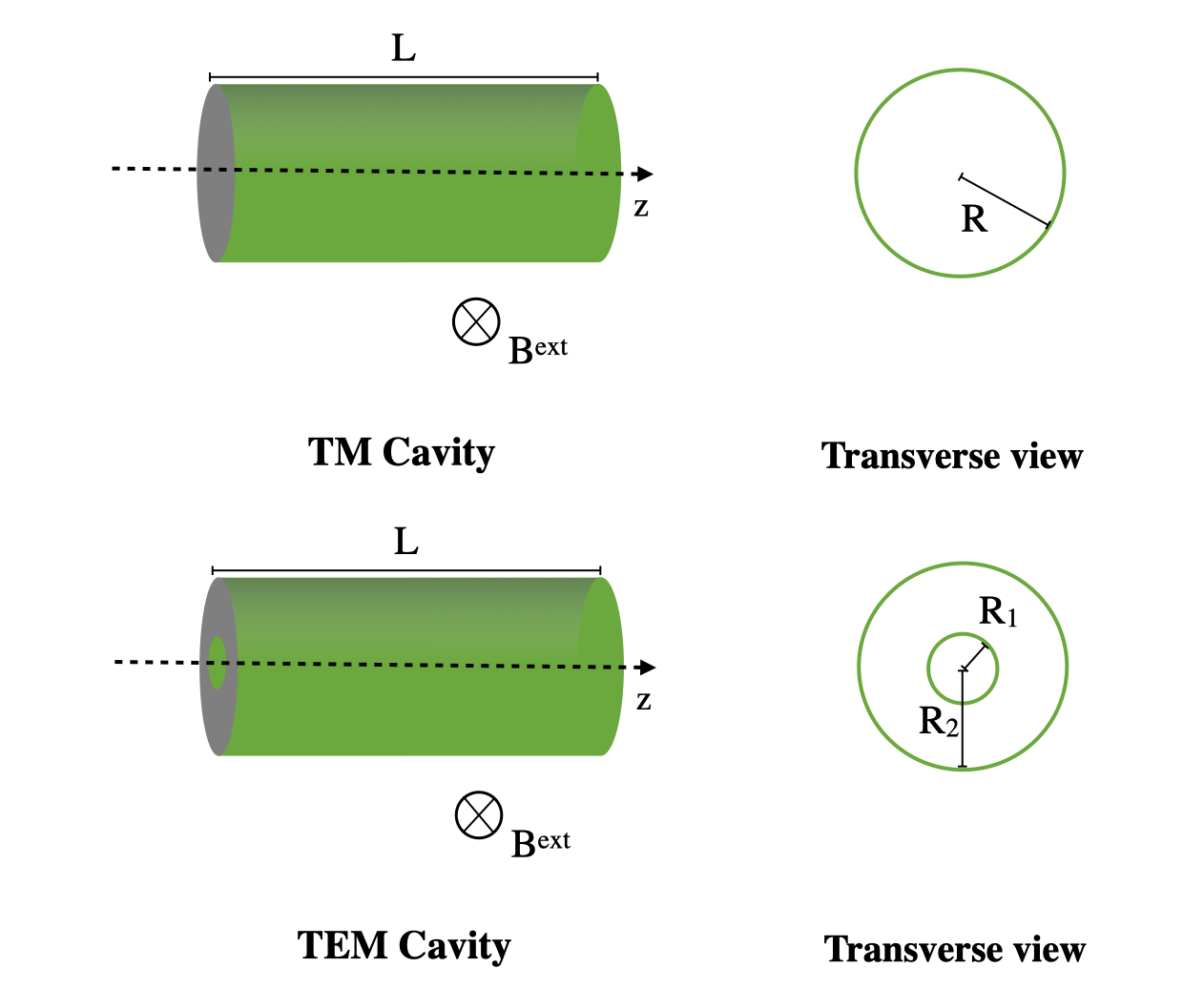}
\caption{Schematic representation of the experimental designs:  a cylindrical TM cavity (top) and TEM waveguide (bottom), into an external static and transverse magnetic field. }
\label{schema}
\end{figure}

Let us now present our proposed experimental set-ups.
 One can either consider the resonance of the induced EM waves inside a cylindrical cavity of radius $R$ or inside
a waveguide made of two (or more) concentric open cylinders with inner radius $R_1$ and outer radius $R_2$.
We denote by $L$ the length of the resonators and by $B^{(0)}_{\rm ext}$ the external magnetic field,  assumed of constant magnitude for simplicity. A schematic representation of our cavities can be found in the figure~\ref{schema}.

We briefly present here the responses of these cavities to an incoming GW signal. The interested reader will find the details of the computations at the end of this paper, in the Appendix.
 In the following, we will assume $c=1$. The starting point is the induction of EM waves when the GW passes perpendicularly to the static magnetic field, as described by Eq. (\ref{lineq}).  Considering the outer magnetic field along the x-direction : $\vec{B^{(0)}_{\rm ext}}=B^{(0)}_{\rm ext}\vec{e_x}\cdot$, we obtain the following wave equation for the induced magnetic field $\vec{B}^{(1)}$
 \be
\left(-\frac{\partial^2}{\pr t^2}+\vec{\Delta}\right) \vec{B}^{(1)}=B^{(0)}_{\rm ext}
\begin{pmatrix}
	\frac{\partial^2  \hp}{\partial z^2} \cos(\phi)+\frac{\partial^2  \hx}{\partial z^2 } \sin(\phi)\\
	-\frac{\partial^2  \hp}{\partial z^2} \sin(\phi)+\frac{\partial^2  \hx}{\partial z^2 } \cos(\phi)\\
	0
\end{pmatrix}
\label{waveB}
\ee
where $\hp$ and $\hx$ are the usual polarizations of the incoming GW in the traceless-transverse gauge. Although there is also an induced electric field $\vec{E}^{(1)}$, the response of the detector is dominated by the induced magnetic field, as we shall see below. 

We can then project this Eq.(\ref{waveB}) on the proper functions of the laplacian operator in cylindrical coordinates. This spectral decomposition is given by
\be
 B^{(1)}_{r,\phi}(t,r,\phi,z)\approx \sum_{k,m,n} \hat{b}^{r,\phi}_{k,m,n}(t) \cdot\psi^{r,\phi}_{kmn}(r,\phi,z)
 \ee
 where $\psi^{r,\phi}_{kmn}(r,\phi,z)$ are the cylindrical harmonics that satisfy the boundary conditions of our EM cavities.
The result of this spectral decomposition allows reducing the above wave equation Eq.(\ref{waveB}) to an ordinary differential equation describing a forced harmonic oscillator for each spectral mode $\hat{b}^{r,\phi}_{k,m,n}$ in our cavity:
 \be
 \frac{d^2\hat{b}^{r,\phi}_{k,m,n}}{dt^2}+\Omega^2_{kn}\hat{b}^{r,\phi}_{k,m,n}=\hat{s}^{r,\phi}_{k,m,n}(t)
 \label{osc}
 \ee
 where $\Omega^2_{kn}$ are the proper frequencies of the resonant cavities and $\hat{s}^{r,\phi}_{k,m,n}(t)$ are the spectral coefficients of the source of the wave equation Eq.(\ref{waveB}).
 
The energy variation $\Delta \mathcal{E}$ inside the cavity is given by, at the leading order (see our Appendix for details), 
\be
\Delta \mathcal{E}\approx \frac{2\pi B_0 }{\mu_0} \cdot \sum_{k} \mathcal{I}_k\hat{b}_{k,1,0}(t)
\label{deltaE}
\ee
where the coefficients $\mathcal{I}_k$ arise from the spatial averaging in the transverse directions (these adimensional quantities depends on the cavity geometry only).  Let us recall that the numbers $k$, $m$ label the transverse 
decomposition in the radial and azimuthal direction, while number $n$ labels the different longitudinal modes. 
We can see that only the $(k,1,0)$ modes, which are constant in the longitudinal direction $z$ (since $n=0$), are contributing to the energy variation at first order in $B^{(1)}$. Since these $(k,1,0)$ modes do not propagate along $z$, one can see that there is no spatial phase shift with this energy variation at first order.
These $(k,1,0)$ modes of the induced magnetic field are sourced by
\be
\hat{s}_{k,1,0}^{r,\phi}(z,t)=\pi B_0 L^2 \mathcal{I}_k\int_{-L/2}^{L/2} \frac{\partial^2  \hp(z,t)}{\partial z^2 }dz
\label{shat}
\ee
The detailed computations from Eq.(\ref{lineq}) to Eqs.(\ref{osc}-\ref{shat}) is available at the end of this paper, in the Appendix.

 A dimensional analysis of Eq.(\ref{deltaE}) leads to the following estimation for
 the order of magnitude of the induced energy variation inside the resonator
 \be
 \Delta \mathcal{E}\approx \frac{2\pi B_0^2 L^3 }{\mu_0}\mathcal{H}_{\rm GW}\mathcal{F}
 \label{estim}
 \ee
 where $\mathcal{H}_{\rm GW}$ is the (dimensionless) amplitude of the strain of the GW and where $\mathcal{F}$ is a adimensional geometrical factor\footnote{In other words, we extract all the dimensional factors in the equation (\ref{deltaE}), letting just one adimensional expression that depends on the geometry of the detector and its frequency sensitivity.} accounting for the shape of the resonator (length and diameter). This factor $\mathcal{F}$ is of the order of unity when the spectrum of the incoming GW matches the resonance bandwidth. In the other cases, we must consider a frequency dependant geometrical factor $\mathcal{F}(\omega)$.  
 
Please note that with our model there is no temporal phase shift in the conversion process. Indeed there is no imaginary part in the Fourier transform of $\Delta E$ and so the complex argument is null. This is mainly due to the fact that Eq.(\ref{osc}) is purely harmonic, without any dissipation, and therefore no phase shift can happen. Instead of a derivation in the time domain, one could also use a frequency approach of the cavities responses to demonstrate this, but this goes beyond the scope of this paper. However, the ohmic losses of energy in the sidewalls of the cavities could be represented by a dissipative term in equation (\ref{osc}), and so a phase shift could appear in the conversion process for resistive cavities. Although ohmic losses will lower the efficiency of the resonance mechanism investigated here, this can be avoided by working with superconducting cavities.

\section{HFGWs from planetary-mass primordial black hole mergers} \label{sec:PBH}

PBH binaries may have formed through two different channels.  First, in the early Universe, when two PBHs form sufficiently close to each other for their dynamics to decouple from the expansion of the Universe, before the matter-radiation equality~\cite{Nakamura:1997sm,Sasaki:2016jop}.  Second, by tidal capture in dense environments~\cite{Bird:2016dcv,Clesse:2016vqa}, such as ultra-faint dwarf galaxies.  In this section, we review the motivations to consider planetary-mass PBHs binaries and we estimate their expected merging rate and gravitational-wave signal, for those two channels.  We then calculate the astrophysical range of resonant electromagnetic detectors as a function of their strain sensitivity.  Finally, we compute for each formation channel the limits that could be set on the abundance of planetary-mass PBHs. 

\subsection{Motivations} 

The progenitor masses and low effective spins of the black hole mergers detected by LIGO/Virgo have revived the interest for PBHs in the $ [1-100] M_\odot$ range~\cite{Bird:2016dcv,Clesse:2016vqa,Sasaki:2016jop,Kashlinsky:2016sdv}.  However it is debated if PBHs could constitute only a small fraction, or up to the totality of the DM in the Universe.  In this context, detecting a sub-solar black hole would almost clearly point to a primordial origin\footnote{See however~\cite{Kouvaris:2018wnh,PhysRevLett.126.141105} for another subsolar black hole formation channel, with different spin predictions, in a specific dark matter scenario.}. Going beyond the simplest but unrealistic assumption of a monochromatic mass function, the distribution of PBHs could span several decades of masses, as it is the case if curvature fluctuations at the origin of PBH formation are nearly scale invariant -- a generic prediction of inflation -- or come from a broad peak in their power spectrum.   Then the known thermal history of the Universe, in particular the QCD transition at $\sim 100$ MeV and the electroweak epoch at $\sim 100$ GeV, should have left imprints in the PBH mass function~\cite{Byrnes:2018clq,Carr:2019kxo}, whatever is the mechanism at the origin of those curvature fluctuations.   These features take the form of a high peak at the solar mass scale and two bumps at $\sim 30 M_\odot$ and $\sim 10^{-5} M_\odot $.   Such an extended mass function could explain a series of puzzling observations  (see~\cite{Clesse:2017bsw,Carr:2019kxo} and references therein) such as unexpected microlensing events, LIGO/Virgo black hole mergers, some properties of ultra-faint dwarf galaxies, unexpected correlations in X-ray and infrared cosmic backgrounds, and super-massive black holes at high redshifts.  The bump in the planetary-mass range is consistent with recent detections of star and quasar microlensing events~\cite{Niikura:2019kqi,Hawkins:2020zie,Bhatiani2019}, which suggest a DM fraction of $f_{\rm PBH} \sim 0.01$ made of compact objects, quite much than one can expect for floating planets, but that could be expected for PBHs in the unified scenario presented in~\cite{Carr:2019kxo}.   Recently, the possible detection of a stochastic GW background at nanoHerz frequencies by NANOGrav~\cite{Arzoumanian:2020vkk} may also hint at the existence of PBHs with planetary~\cite{Domenech:2020ers} or stellar~\cite{DeLuca:2020agl,Vaskonen:2020lbd,Kohri:2020qqd} masses, and Wang et al.~\cite{Wang2019,Wang2020} puts some constraints on the PBH abundance for the current detectors to probe a stochastic GW background made of PBHs.   However all these observations could have another origin and the derived limits are still subject to large astrophysical uncertainties.   In the future, it is therefore important to find complementary ways to probe the existence of such objects, and to distinguish their nature and origin. 

As we show in this paper, HFGW detectors will have the ability to detect or set new limits on the abundance of light, subsolar PBHs, of mass $m_{\rm PBH} \sim 10^{-5} M_\odot$.   HFGWs are indeed produced during the merging phase of such light PBHs.   The frequency associated to the innermost stable circular orbit (ISCO), when the GW emission is close to maximal, is given by 
\be
f_{\rm ISCO} = \frac{ 4400\, {\rm Hz} }  {(m_1 + m_2) / M_\odot} ~.
\label{fisco}
\ee
with $m_1$ and $m_2$ the masses of the two binary components.  A frequency of $200$ MHz thus corresponds to a PBH mass of $10^{-5} M_\odot$, the same order than the mass of the lenses at the origin of the microlensing events reported in~\cite{Niikura:2019kqi}.  Nevertheless, for being an interested HFGW signal, one needs to investigate if the merging rate of such PBHs can lead to at least $\mathcal O(1) $ mergers per year within the HFGW detector range.  

PBHs therefore constitute a target of much interest for our experimental concept of EM detection of HFGWs. From the amplitude and spectral response of the resonant detectors, we will characterize the expected signals from PBHs mergers for a large range of progenitor masses in the interval $[10^{-8};10^{-3}] \, M_\odot$, located at $1$ Gpc distance.  Then, for a given detector sensitivity, we will compute the expected limits on the PBH abundance.


\subsection{Gravitational-waves from inspiraling binaries}

A good estimation of the GW strain produced at a given frequency $f_{\rm GW}$ during the inspiralling phase of a black hole binary is provided by the Post-Newtonian approximation~\cite{Antelis:2018sfj},
\be \label{eq:strainPBHs}
h \approx \frac{2}{D} \left( \frac{G \mathcal M }{c^2}  \right)^{5/3} \left( \frac{ \pi f_{\rm GW}}{c} \right)^{2/3} ~,
\ee
where $\mathcal M \equiv (m_1  m_2)^{3/5}/(m_1+m_2)^{1/5} $ is the binary chirp mass and $D$ is the distance to the observer.
The GW emission is close to maximal at the ISCO frequency and,
for a given chirp mass, for equal mass binaries, $m_1 = m_2 = m_{\rm PBH}$.   For an experiment with a detector strain sensitivity $h_{\rm det}$ at this maximal frequency,  the corresponding astrophysical reach $D_{\rm max}$ is given by
\be  \label{eq:Dmax}
D_{\rm max} \approx 1.6 \times  \frac{(m_{\rm PBH}/M_\odot) }{h_{\rm det} \times 10^{20}} {\rm Mpc}.
\ee
For instance, for a strain sensitivity of $h_{\rm det} \sim 10^{-25}$ and $m_{\rm PBH} \sim 10^{-5} M_\odot$, eventual mergers towards the galactic center, in the Milky Way DM halo, or in satellite ultra-faint dwarf galaxies, could be detected.   For a better sensitivity down to $h_{\rm det} \sim 10^{-30}$, corresponding to the optimal sensitivity of the proposed designs of resonant  EM detectors, then one would probe planetary-mass PBH mergers in more distant galaxies.

\subsection{Merging rate of primordial binaries}

If PBHs are spatially randomly distributed at formation, it happens that two PBHs form so close to each other that their gravitational attraction overpasses the effect of the Hubble-Lema\^itre expansion at some point before matter radiation equality.  In such a case, they directly form a binary whose orbital parameters and lifetime do not only depend on the two black hole masses but also on the mass and distance of the nearest PBHs.  Eventually, it takes of the order of the  age of the Universe for the PBH binary to merge.   The merging rates $\tau $ today associated with this binary formation channel and an arbitrary mass function have been evaluated in ~\cite{Kocsis_2018,Raidal:2018bbj,Gow:2019pok,Liu2019} as 
\bea
       R^{\rm prim} (m_1, m_2) & \equiv & \frac{{\rm d} \tau}{{\rm d} \ln m_1 {\rm d} \ln m_2}   \nonumber  \\ 
       & \approx & \frac{1.6 \times 10^6}{\rm Gpc^3 yr} f_{\rm PBH}^2 f(m_1) f(m_2) f_{\rm sup}  \nonumber  \\ 
       & \times & \left(\frac{m_1 + m_2}{M_\odot}\right)^{-\frac{32}{37}} \left[\frac{m_1 m_2}{(m_1+m_2)^2}\right]^{-{\frac{34}{37}}}
\eea
where $f(m)$ is the today density distribution of PBHs normalized to one ($\int f(m) {\rm d} \ln m = 1$) and $f_{\rm PBH}$ is the integrated DM fraction made of PBHs.  We also define an effective parameter
\be
\tilde f_{\rm PBH} (m_{\rm PBH}) \equiv f_{\rm PBH} f(m_{\rm PBH}) f_{\rm sup}^{1/2}
\ee 
that includes a rate suppression factor ($f_{\rm sup}$) to take into account the possible rate suppression due to binary disruption by early-forming clusters, an effect put in evidence by N-body simulations when $f_{\rm PBH} \gtrsim 0.1$~\cite{Vaskonen:2019jpv}.  In such a case, one can recover the LIGO/Virgo merging rates inferred from the recent detections of GW190425, GW190521 and GW190814 involving at least one BH in the mass gaps, with $f_{\rm PBH} = 1$ and $f_{\rm sup} \simeq 0.0025$~\cite{Clesse:2020ghq}.  In the opposite case, $f_{\rm sup}= 1$ and $\tilde f_{\rm PBH}$ simply represents the DM density fraction made of PBHs at a given mass and within a unit logarithmic mass interval.  

If one considers the merging rates of equal-mass binaries that produce the largest strain signal, one gets
\bea
    R^{\rm prim}(m_{\rm PBH}) &\approx&  \frac{3.1 \times 10^6}{\rm Gpc^{3} yr} \tilde f_{\rm PBH}^2 
    \left( \frac{m_{\rm PBH}}{M_\odot}\right)^{-0.86}.
 \eea
  In turn, one can determine the radius of the sphere in which one expects one event per year,
  \be
  D^{\rm prim}_1 = \left( \frac{4\pi}{3 } R^{\rm prim}\right)^{-1/3} \approx 4.2 \, {\rm Mpc} \times \tilde f_{\rm PBH}^{-2/3} \left( \frac{m_{\rm PBH}}{M_\odot}\right)^{0.29}~.
  \ee
  For simplicity we neglected the effects of redshift that are anyway insignificant for most of the considered cases.  For instance, in the scenario of~\cite{Carr:2019kxo,Clesse:2020ghq} with $f_{\rm PBH} =1$, $f_{\rm sup}=0.0025$ and $f(10^{-5} M_\odot) \simeq 10^{-2}$, one gets $D_1^{\rm prim} (10^{-5} M_\odot) \approx 23 \, {\rm Mpc}$.
  Using Eqs.~(\ref{eq:strainPBHs}) and~(\ref{eq:Dmax}), one then obtains the required GW strain sensitivity to detect one of these merger events per year, 
    \bea
    h_{1}^{\rm prim}  & \approx & 3.8 \times 10^{-21}
    \tilde f_{\rm PBH} ^{2/3} \left(\frac{m_{\rm PBH}}{M_\odot}\right)^{0.7}  \\
    & \approx & 8.3 \times 10^{-19}
    \tilde f_{\rm PBH} ^{2/3}  \left( \frac{\rm Hz}{f} \right)^{0.7} ~,
    \eea
  which can be typically targeted by GW experiments operating at  frequencies from kHz up to GHz.   This relation can be inverted to obtain a limit on the DM fraction at a given mass (if $f_{\rm PBH} <0.1$) in case of null detection, as a function of the strain sensitivity,
  \bea
 \tilde f_{\rm PBH}  \lesssim 9.1 
 \left[  \frac{h_{\rm det}}{10^{-20}}\right]^{3/2} \left( \frac{m_{\rm PBH}}{M_\odot} \right)^{-1.07} .
  \eea
However, the strain sensitivity of the detector depends on the waveform and signal duration, which depend on the PBH mass.  It is thus more adequate to compute a limit on the PBH abundance taking these effects into account and instead assuming an EM power sensitivity, which we do in the next section.
For instance, we obtain with the two proposed experimental designs and a power sensitivty of $10^{-10}$W, limits that are competitive with the current microlensing limits at the same mass scale.  These are represented in our final Figure~\ref{fig_fDM}.


Finally, we point out that when our analysis was being finalized, the authors of~\cite{Boehm:2020jwd} have claimed that the rates from primordial binaires are highly suppressed compared to previous calculation.  The reason is a subtle general relativistic effect arising when one considers geodesics in black hole exterior spacetime metrics that are FLRW asymptotic.   If this claim is correct, primordial binaires are by far outside the reach of EM detectors, but one can nevertheless consider the merging rates inside PBH clusters, which we detail thereafter.

\subsection{Merging rate from PBH clusters}

The second binary formation channel is through dynamical capture in dense PBH halos.  As any other DM candidate, PBHs are expected to form halos during the cosmic history, and their clustering properties determine the overall merging rate.   For instance, for a monochromatic mass spectrum and a standard Press-Schechter halo mass function, 
    one gets a rate~\cite{Bird:2016dcv} 
    \be \label{eq:capturerate}
  R^{\rm capt}  \sim f_{\rm PBH}^2 \times \mathcal O(1-100) \, {\rm yr^{-1} \, Mpc}^{-3}
    \ee
 that is independent of the PBH mass.  For more realistic extended mass functions, the abundance, size and evolution of DM halos, partially or entirely made of PBHs, is impacted by several effects, see e.g.~\cite{Chisholm:2005vm,Chisholm:2011kn,Belotsky:2018wph,Carr:2018rid,Suyama:2019cst,MoradinezhadDizgah:2019wjf,Matsubara:2019qzv,Young:2019gfc,Padilla:2020xlo,DeLuca:2020jug,Jedamzik:2020ypm}.  Let us mention a Poissonian noise from the discrete nature of PBHs, a seeding effect from heavy PBHs, the enhancement of the primordial power spectrum at the origin of PBH formation, the dynamical heating and evaporation of clusters, etc.   These effects can  either boost or suppress the merging rates from clusters and make the whole clustering dynamics a rather complex and model-dependent process, subject to large uncertainties.  Invoking clustering is also crucial to evade microlensing limits on stellar-masses~\cite{Garcia-Bellido:2017xvr,Calcino:2018mwh} in scenarios with $f_{\rm PBH}=1$.  As an alternative of using uncertain theoretical predictions, on can instead infer an upper limit on the PBH merging rate from LIGO/Virgo observations, see e.g~\cite{Clesse:2020ghq}.  
The merging rate from tidal capture in PBH clusters for an arbitary mass function is given by~\cite{Clesse:2016vqa,Clesse:2020ghq}
\bea 
    R^{\rm capt} (m_1,m_2) &\equiv&  \frac{{\rm d}\tau}{{\rm d} \ln m_1 {\rm d} \ln m_2 } \nonumber \\
    &\approx& R_{\rm clust}  f_{\rm PBH}^2 \times f(m_1) f(m_2)  \nonumber \\
    & \times & \frac{(m_1 + m_2)^{10/7}}{(m_1 m_2)^{5/7}} \rm{yr^{-1}Gpc^{-3}}, 
     \label{eq:ratescatpure2}
  \eea
where $ R_{\rm clust} $ is an effective parameter encompassing the clustering properties.  For $f_{\rm PBH} =1$ and $R_{\rm clust} \approx 450$, these rates are consistent with the latest LIGO/Virgo observations, for a broad PBH mass function impacted by the transient reduction of the critical threshold of PBH formation at the QCD epoch.  As already mentioned, this effect is unavoidable and may have induced a peak around $2.5 M_\odot$ and a bump around $30 \, M_\odot$ in the PBH mass function.   In this scenario, around one percent of the DM could be made of planetary-mass PBHs around $10^{-5} M_\odot$.  Like for primordial binaries, we define an effective parameter
\be
\tilde f_{\rm PBH} \equiv \left( \frac{R_{\rm clust}}{450}\right) \times f_{\rm PBH} f(m_{\rm PBH})
\ee
representing the DM density fraction at a given mass and per logarithmic mass interval, in the above mentioned scenario.
One then obtains the merging rate for equal-mass binaries, 
\be
R^{\rm capt} (m_{\rm PBH}) = 1.2 \times 10^3 \tilde f_{\rm PBH}^2~,
\ee
the corresponding source distance $D_1^{\rm capt} $,
\be
D_1^{\rm capt} \approx 58 \, {\rm Mpc} \times \tilde f_{\rm PBH}^{-2/3},
\ee
and the required experimental strain sensitivity to detect one event per year,
    \be
    h_{\rm 1}^{\rm capt} \approx  2.7 \times 10^{-22}  \tilde f_{\rm PBH}^{2/3} \left( \frac{m_{\rm PBH}}{M_\odot}\right) .
    \ee   
     This GW signal is therefore typically lower than for PBH binaires on planetary-mass scales.

Finally, like for the case of primordial binaries, we have derived the expected limits on $\tilde f_{\rm PBH}$ for a given experimental strain sensitivity,
  \bea
 \tilde f_{\rm PBH}  \lesssim 2.5 \times 10^{4} 
 \left[  \frac{h_{\rm det}}{10^{-20}}\right]^{3/2} \left( \frac{m_{\rm PBH}}{M_\odot} \right)^{-3/2}.
  \eea
 As an example, if $h_{\rm det} = 10^{-30}$ and $m_{\rm PBH} = 10^{-5}$, one gets $\tilde f_{\rm PBH} \lesssim 7 \times 10^{-4}$, which is better than the current microlensing limits.   Again, it is more accurate to assume a power sensitivity rather than a strain sensitivity.  Doing so, the corresponding limits on $\tilde f_{\rm PBH}$ have been represented in our final Figure~\ref{fig_fDM}.
 


\section{Probing PBH mergers with resonant EM detectors of HFGWs}\label{sec:det-PBH}

\begin{figure}[ht!]
\centering
\includegraphics[clip,scale=0.26]{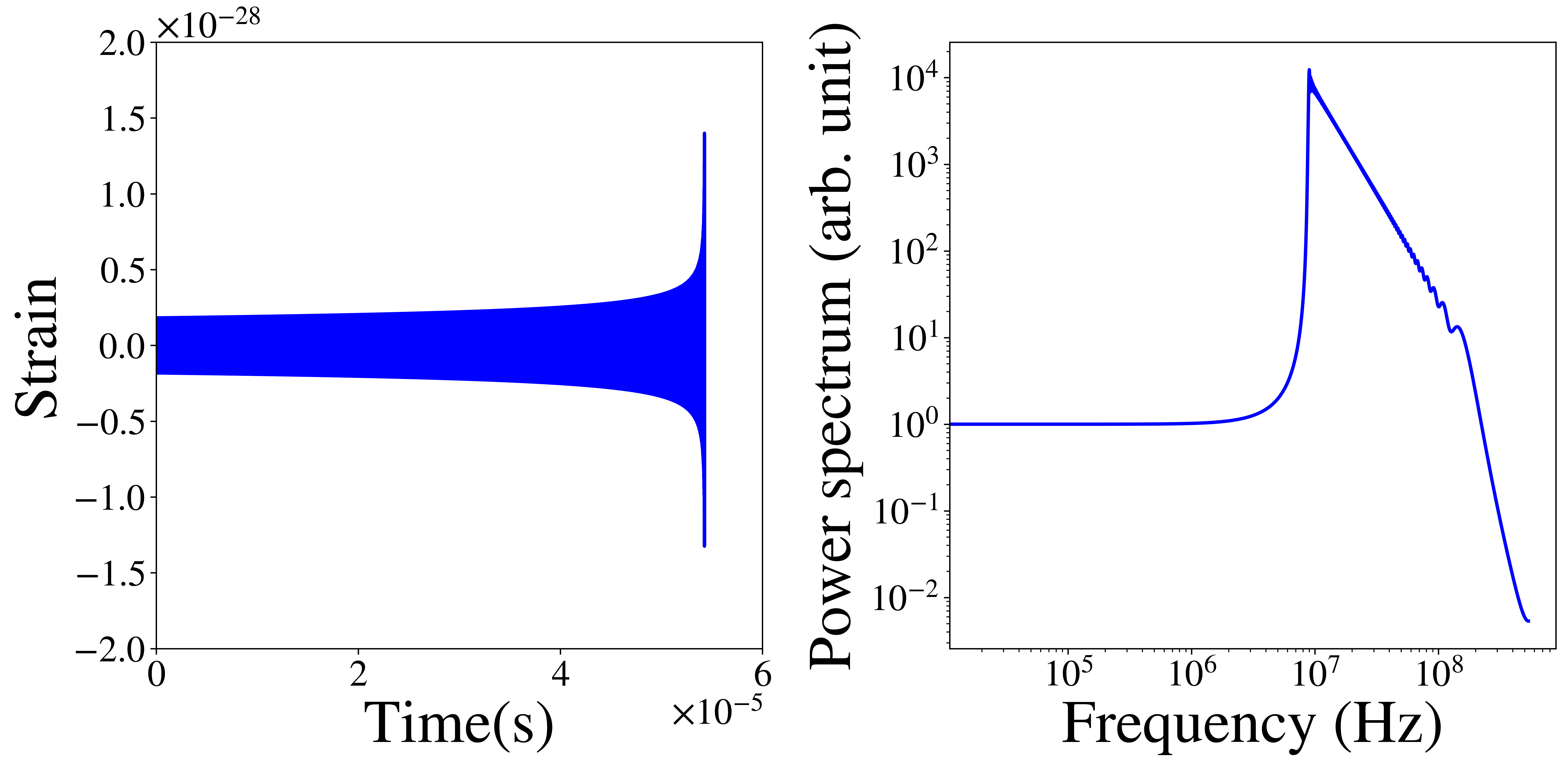}
\caption{Left: GW waveform $h_+$ for the final inspiraling phase of a PBH binary with component masses of $10^{-5} M_\odot$, at a distance of $1\,\rm Gpc$, in the post-newtonian (4PN) approximation.  Right: corresponding GW signal power spectrum. }
\label{fig1}
\end{figure}

Let us consider GW trains produced during the final inspiraling phase of PBHs binaries of different masses and passing through the resonator 
and let us analyze the induced
EM radiation.  For simplicity, we consider a GW propagation colinear 
with the longitudinal
axis of the EM resonator and perpendicular to the outer magnetic field. In this case, the GW can be approximated by a plane wave, i.e. $h_{+,\times}=h_{+,\times}(z,t)$ since the radius of the detector is much smaller than the incoming wavefront for a distant source.  Some inclination of the direction of the incoming GW would result in a signal of lower amplitude, since only the component of the GW in the direction of the outer magnetic field contributes to the inverse Gertsenshtein effect.  
Yet, this ideal case allows to illustrate the physical process and to estimate 
the expected response of the detector and the output signal.  
We have computed the 
typical EM signals 
for resonators of various shapes.  
Post-Newtonian time-domain GW waveforms are generated by using 
the \texttt{LALSuite} library \cite{lalsuite}, assuming a 4PN approximation. 
For all the simulations, we choose a signal sampling frequency that is four time the 
ISCO frequency [Eq.~(\ref{fisco})].  
The initial frequency of the signal is set to $f_{\rm ISCO}/25$.   On Fig.~\ref{fig1}, we provide an example of the GW waveform produced by the merging of a PBH binary with component masses of $10^{-5} M_\odot$ and located at a distance of $1\,\rm Gpc$. The amplitude of the GW strain at reception, denoted $\mathcal{H}_{\rm GW}$, is of order $10^{-28}$ and the signal duration is of order
$10^{-5}s\cdot$. 

\begin{figure}[ht!]
	\begin{tabular}{c}
\includegraphics[clip,scale=0.26]{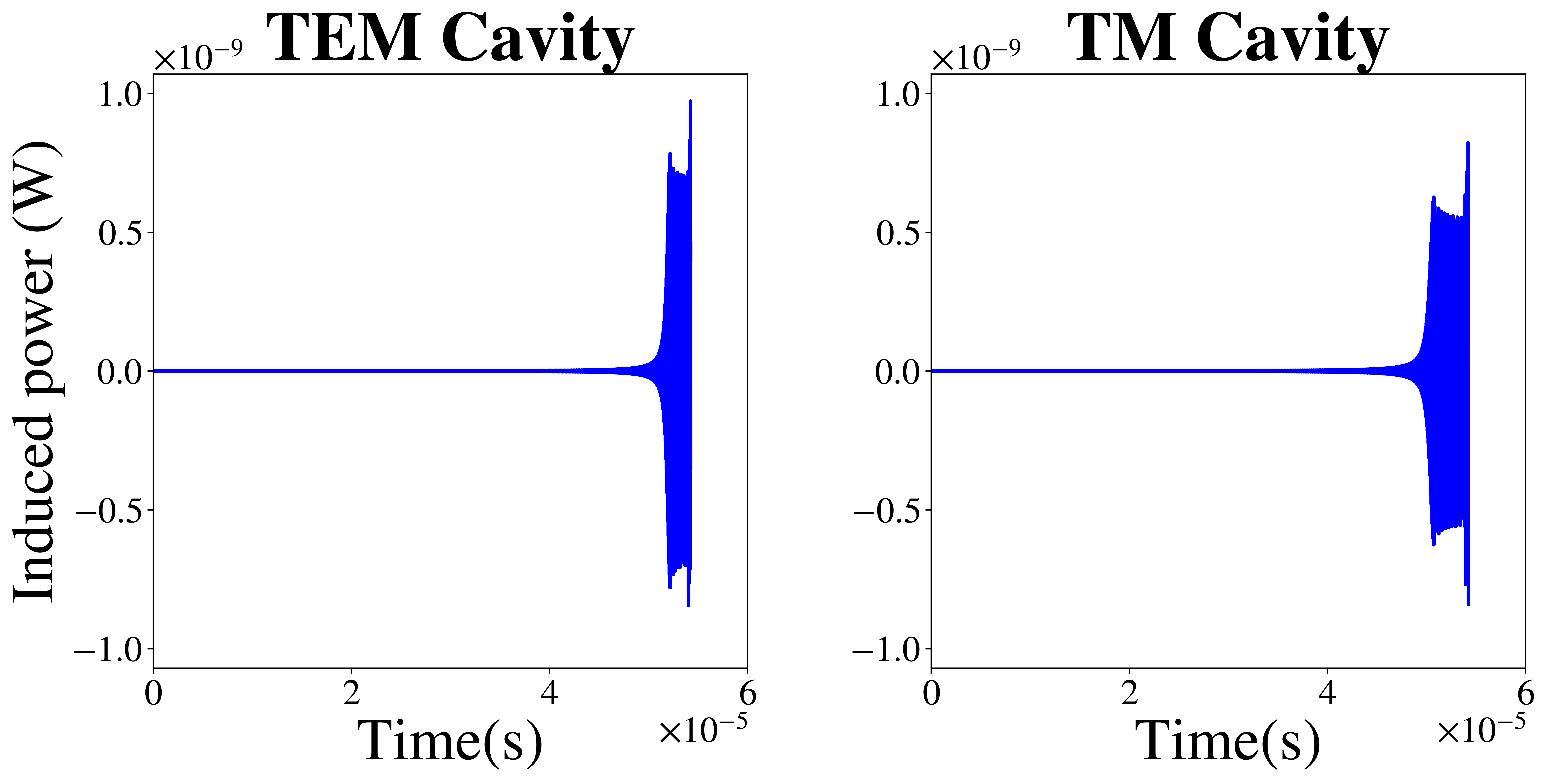}\\
\includegraphics[clip,scale=0.26]{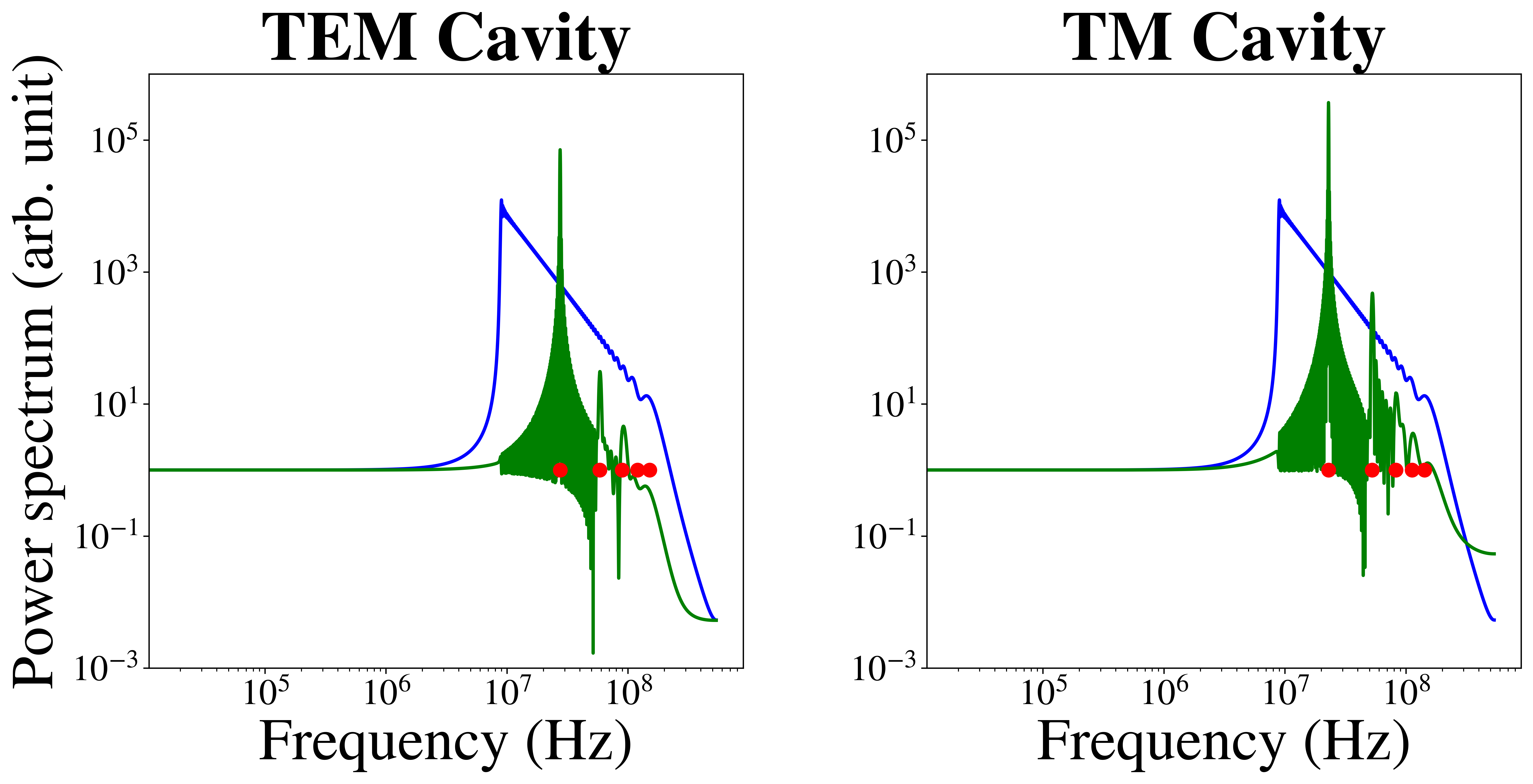}

	\end{tabular}
	\caption{Top: time evolution of the induced EM power in the two resonators.  Bottom:  GW signal power spectrum (blue) and induced power spectrum (green). The detectors are 1-meter long, in a 5\,T constant magnetic field. The outer radius is 5\,m and for TEM coaxial case, with an inner radius of 10\,cm.  
	The peaks in the power spectrum correspond to the resonance frequencies, represented by the red dots. 
	}
	\label{figres}
\end{figure} 

Then, by using Eq.~(\ref{deltaE}) and solving the forced harmonic oscillator equations, Eqs.~(\ref{osc}), we compute the induced EM power for the two designs of resonators. 
Our results are displayed on Fig. \ref{figres}, for 1-meter long resonators in a 5\,T constant magnetic field. The radius of the resonators is set to 5\,m and for the TEM coaxial design, we consider a 10 cm inner radius.  As expected, the induced power increases near the merger.  In each case, we have also computed the frequency spectrum of the induced power, which exhibits peaks
corresponding to the resonance frequencies, coming from Eqs.~(\ref{freq1}) and (\ref{freq2}) in the Appendix. The power rms values are $1.00\times 10^{-10}$ W for the TEM resonator and  $1.03 \times 10^{-10}$~W for the TM resonator.

\begin{figure}[ht!]
\includegraphics[clip,scale=0.45]{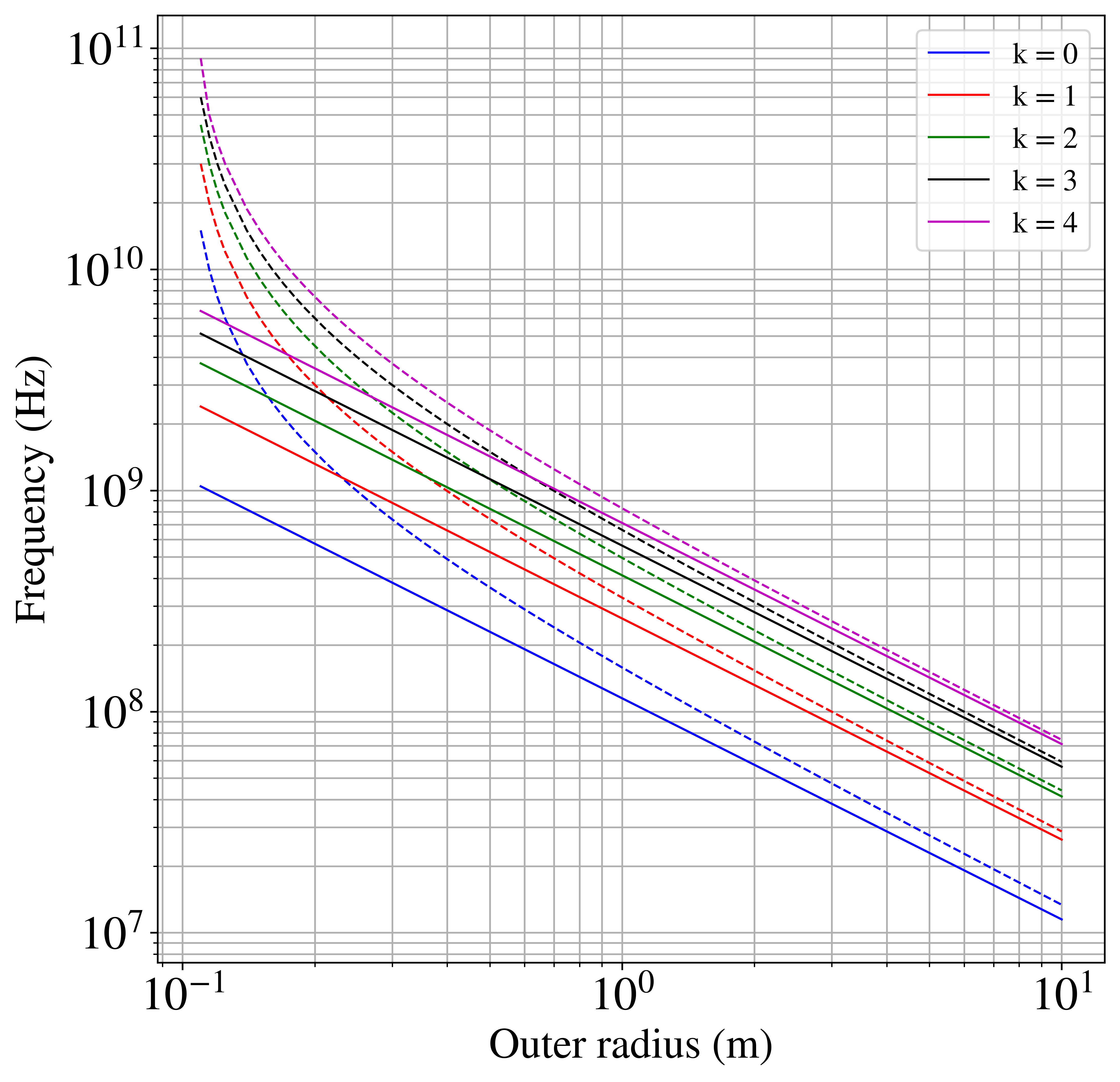}
\caption{The first five proper frequencies of the cavity (TM resonator, straight lines) and the waveguide (TEM resonator, dashed line) as a function of their outer radius (assuming a length of $L=1$\,m and an inner radius of $R_1=0.1$\,m for the waveguide). 
}
\label{figgeo}
\end{figure}

\begin{figure}[ht!]
	\centering
	\begin{tabular}{c}
	\includegraphics[clip,scale=0.26]{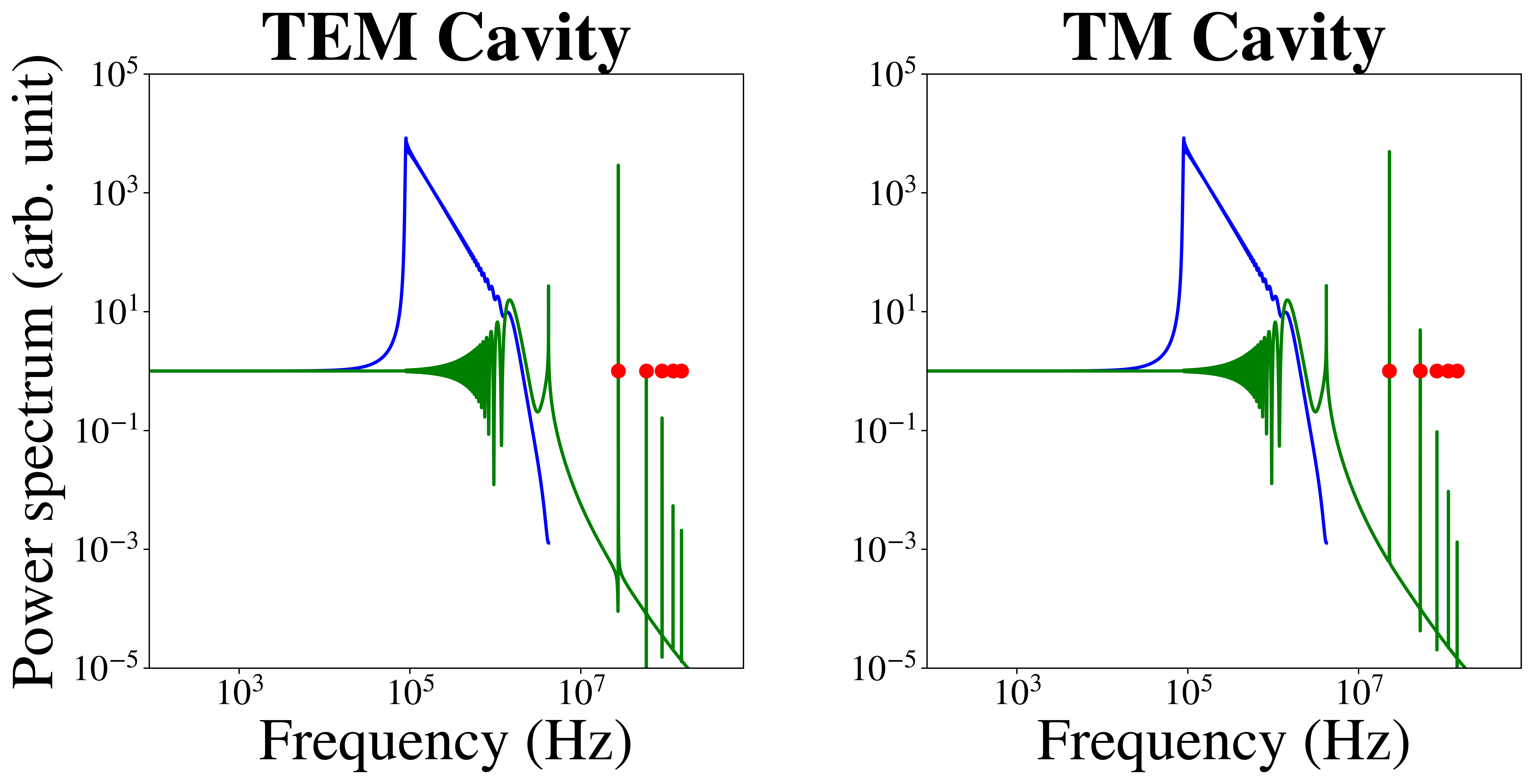}\\
		\includegraphics[clip,scale=0.26]{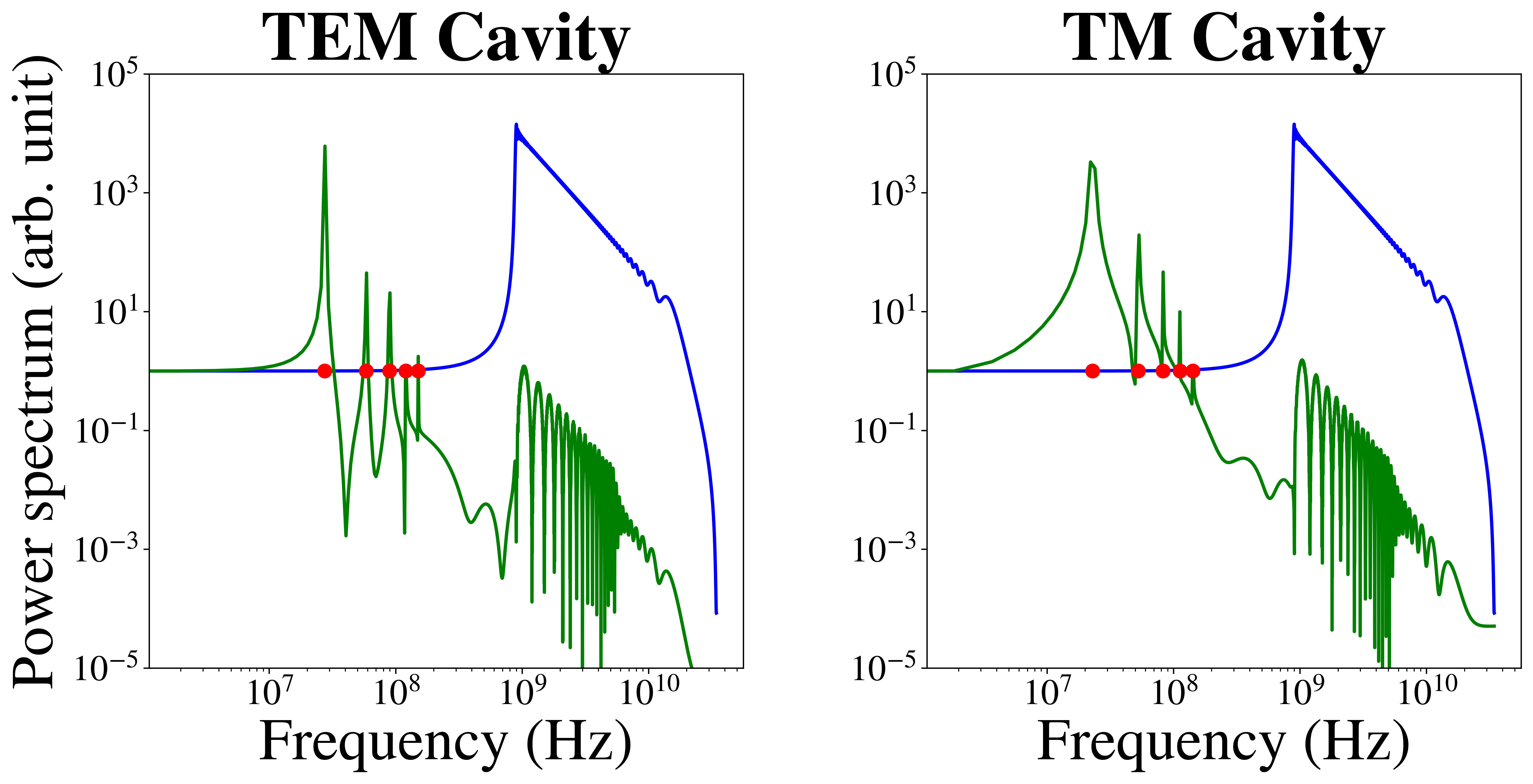}\\
	\end{tabular}
	\caption{Same as the bottom panel of Fig.~\ref{figres}, but for a PBH mass of $10^{-3}\, M_\odot$ (top pannel) and $10^{-7}\, M_\odot$ (bottom panel).  In the first/second case, the GW frequency is below/above the resonance frequencies (represented by the red points) of the detector.
	}
	\label{figres2}
\end{figure}

Fig.~\ref{figgeo} shows the first five proper frequencies of the TEM and TM cavities as a function of their outer radius, within the range $10$\,cm to $10$\,m, for a detector of one meter length.  For the TM resonator, these frequencies range from 10\,MHz to GHz,
while the TEM waveguide can reach higher frequencies for the same detector size.
In the case of an outer radius almost equal to the inner one (thin case), the resonance frequencies 
can be one order of magnitude larger than for a TM resonator. 
We point out that it would even be possible to use several concentric TEM waveguides to further extend the range of resonance frequencies. Fig~\ref{figgeo} therefore illustrates how one could play with the geometry of the experimental set-up in order to optimize the detector response to some motivated GW signal. 
The PBH mass is another parameter to consider.  For the same detectors as in Fig.~\ref{figres}, we have simulated the detection of the signal for two additional PBH masses, of 
$10^{-3}$ and $10^{-7} M_\odot$, the corresponding power spectra are shown in Fig.~\ref{figres2}.
For $10^{-3}\, M_\odot$, the resonance frequencies are higher than the ISCO frequency.  The resulting EM power spectrum in the detector exhibits a continuous set of frequencies that matches the incoming GW power spectrum followed by the excitation of the detector resonance frequencies that constitutes the highest values in the induced radiation power spectrum. The continuous set of induced frequencies comes from the particular solution of the forced oscillator Eq. (\ref{osc}), while the excitation of the detector resonance frequencies comes from the general solution.  

In the $10^{-7}\, M_\odot$ case, the detector resonance frequencies are lower than the main GW signal frequencies. Therefore we have a clear separation between the EM power spectrum that is induced by the resonance frequencies of the detector and the one induced by a range of frequencies inherited from the incoming GW signal. In Fig.~\ref{figres2}, the part of the power spectrum induced by the resonance frequencies is overestimated. Indeed the numerical FFT of the incoming GW leads to a plateau at low frequencies as shown in Fig.~\ref{fig1}. As a result this plateau will be excited by the proper frequencies of the cavity leading to an overestimation of this part of the induced power spectrum. We show below how to avoid  this numerical limitation.


\begin{table}[hb]
	\centering
\begin{adjustbox}{width=\columnwidth,center} 
\begin{tabular}{|c|c|c|c|c|c|} 
	\hline 
	$m_{\text{PBH}}$ ($M_\odot$) & time (s) & $\mathcal{H}_{\rm GW}$ & $f_{\text{ISCO}}$ (Hz) & $P_{\text{RMS}}$ TEM (W) & $P_{\text{RMS}}$ TM (W) \\ \hline  
$10^{-3}$ & 5.43e-03 & 1.15e-26 & 2.20e+06 & 2.37e-14 & 3.19e-14  \\ \hline
$10^{-4}$ & 5.43e-04 & 1.37e-27 & 2.20e+07 & 2.98e-12 & 4.96e-12  \\ \hline
$10^{-5}$ & 5.43e-05 & 1.40e-28 & 2.20e+08 & 1.00e-10 & 1.03e-10  \\ \hline
$10^{-6}$ & 5.43e-06 & 1.17e-29 & 2.20e+09 & 1.51e-11 & 6.31e-12  \\ \hline

\end{tabular}  
\end{adjustbox}
\caption{GW signal duration, maximal strain, ISCO frequency and the corresponding induced power in the resonant cavity or waveguide, for different values of the PBH mass.
The detectors are 1~m long for an outer radius of 5~m.  The inner radius in the TEM case is 10\,cm.  The transverse static magnetic field is set to $5$~T.  The minimal frequency of the GW waveform signal is set to $f_{\rm ISCO} /25$. The bandwidth of the resonant frequencies considered is $\left[2\cdot10^{7},2\cdot10^{8}\right]$ Hz  }
\label{table}
\end{table}

\begin{figure*}
 	\includegraphics[clip,scale=0.42]{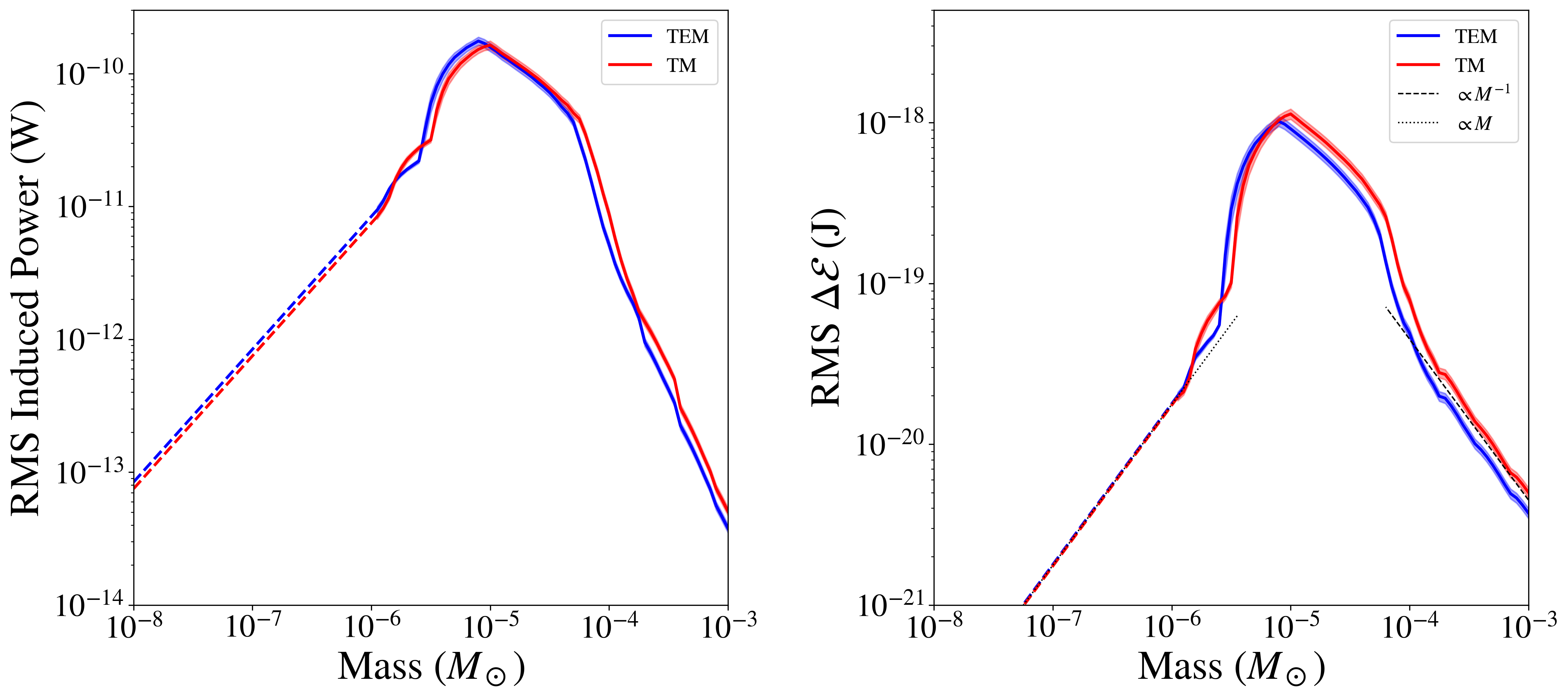}
 	\caption{RMS values of the induced power (left) and energy variation (right) in the TEM (blue) and TM (red) detectors as in Table~\ref{table}, as a function of the PBH mass, for mergers located at a distance of one Gpc.
 	The excitation of resonant frequencies boosts the signal in the range between $10^{-6}$ and $10^{-4} M_\odot$. Below $10^{-6}M_{\odot}$ we plot in dashed lines the expected behaviour due to numerical limitation in our simulations. Asymptotic behaviour coming from our analysis has been plotted in black dotted lines.
  	}
  	\label{figmass}
  \end{figure*}

Finally, we have computed the power in a 1-meter long resonator of 5-meter radius in a 5T magnetic field, induced by the merging of PBHs with a broad range of masses.  Our results are summarized in Table~\ref{table}.  We find that the induced power is maximal for a PBH mass around $10^{-5}\, M_{\odot}$, i.e. when the corresponding GW frequency lies within the range of the resonance frequencies.
We have shown in Fig. \ref{figmass} the rms power as a function of the PBH mass, for the same experimental configuration.  
In order to do so, we have considered 41 simulated signals of PBH mergers for 61 different PBH masses (i.e., 41 signals per discretization point in mass, and each of these signals corresponding to different initial frequency ranging from $f_{\rm ISCO}/30$ to $f_{\rm ISCO}/10$).  Because of the numerical problem at masses beyond $10^{-6}M_{\odot}$(poor resolution of the FFT of the GW signal at low frequencies), we extrapolate our data with the expected behaviour that we will develop further in this section.
When the GW spectrum covers the fundamental resonance modes of the detector, its response is maximal and this occurs for PBH masses covering two orders of magnitude around $10^{-5}M_{\odot}$.

In addition, we have also shown on Fig.~\ref{figmass} the rms energy variation inside the cavity, as a function of the PBH mass.  The induced energy typically corresponds to millions of photons. It is well above the sensitivity of the ADMX experiment, which has achieved an effective instrumental noise temperature of order $\mathcal O(1)$K at similar frequencies~\cite{ADMX,ADMX2} and with a similar value of the magnetic field.  Further work is however needed in order to quantify more accurately the energy or power sensitivity that could be achieved with a similar technology.

For masses larger than $10^{-4}\, M_\odot$, the frequencies associated to the GWs are smaller than the proper frequencies of the cavity and so the amplitude of the total solution of (\ref{osc}) is proportional to the amplitude of the source term, divided by the square of the proper frequencies of the cavity.  
So the behavior of the particular solution is the same as the source term. From Eq.(\ref{waveB}), we know that the source term is proportional to the second time derivative of the strain. On one hand, the GW strain is proportional to PBH mass (see  Eq.(\ref{eq:strainPBHs})). On the other hand, the maximal GW frequency is inversely proportional to the PBH mass. As a result, the energy released in the detector goes like $1/m_{\rm PBH}$, which explains the linear decreases observed at large mass.

As mentionned earlier, because of a numerical limitation we do not plot the induced energy and RMS power for masses smaller than $10^{-6}\, M_\odot$. Nonetheless we can extrapolate the expected behaviour of the mass dependence. The analysis is the same as in the large mass regime except that this time it is the proper frequencies of the cavity that are smaller than the frequencies associated to the GWs. Thus the amplitude of the total solution of (\ref{osc}) is now proportional to the amplitude of the source term, divided by the square of the GW frequencies. As these frequencies are inversely proportional to the PBH mass, the behaviour of the energy released in the detector is $m_{\rm PBH}^2$ times what it was in the previous case. So the energy variation goes like $m_{\rm PBH}$. We will show this in a forthcoming paper by using a frequency domain approach.

Finally, we observe that there is no significant difference between the TEM and the TM detectors, highlighting some robustness with respect to the choice of the experimental set-up.

\begin{figure}[htb!]
\includegraphics[clip,scale=0.6]{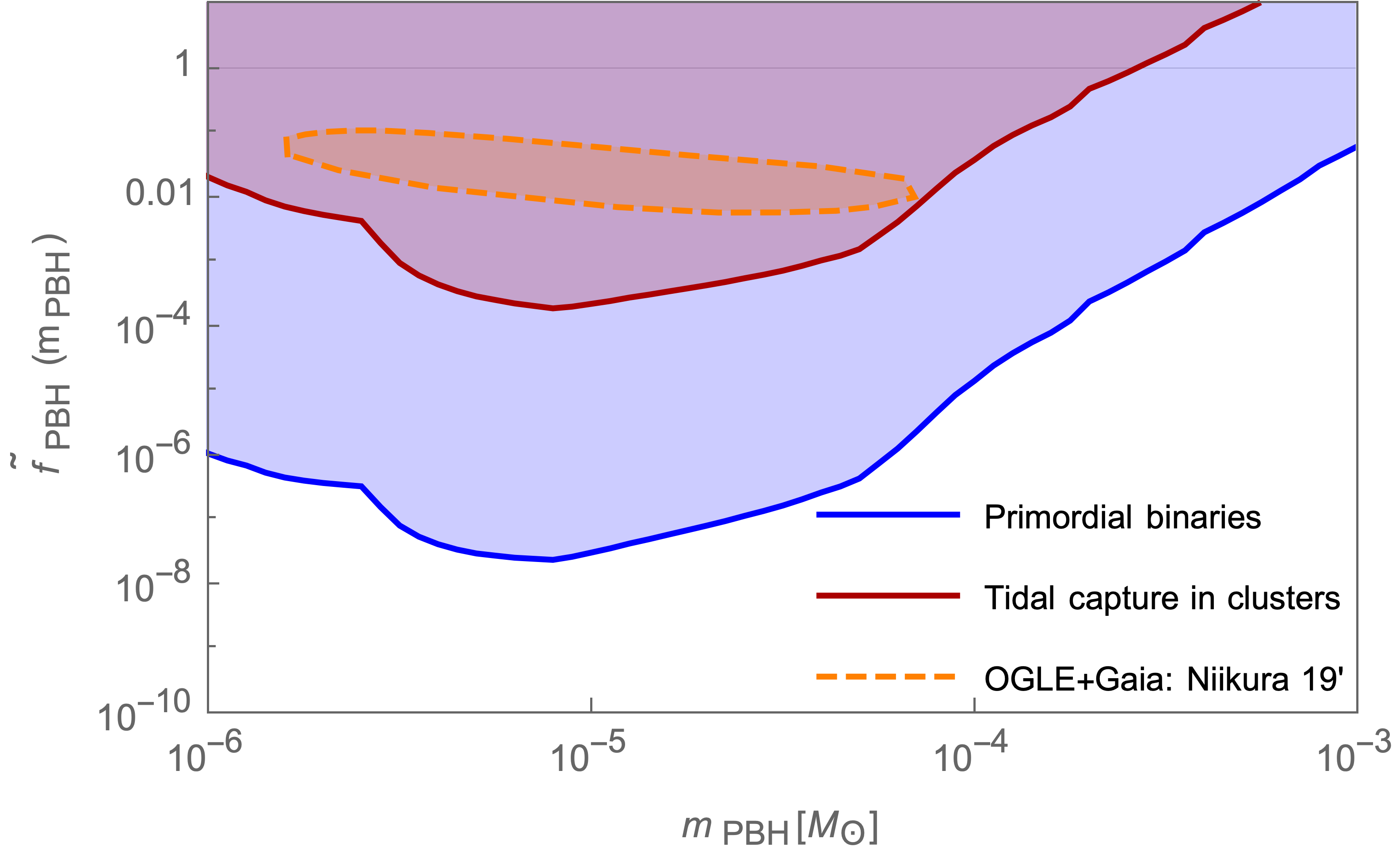}
\caption{Expected limits on the effective parameter $\tilde f_{\rm PBH}$ corresponding to the dark matter density fraction in PBHs at a given mass and per logarithmic mass interval in the two models described in the text: primordial binaries (blue limit) and tidal capture in halos (red limit).  These limits are computed for the proposed experimental designs of EM resonant HFGW detectors, assuming a power sensitivity of $10^{-10}$W, achievable with current technology.  The orange curve shows the possible abundance of planetary-mass PBHs inferred form recent microlensing observations towards the galactic bulge~\cite{Niikura:2019kqi}.}
\label{fig_fDM}
\end{figure}

As a final step, we have combined our calculations of the PBH merging rates and of the induced power as a function of the PBH mass for mergers at a fixed distance of 1 Gpc, in order to forecast limits on the PBH abundance for a fixed detector sensitivity of $10^{-10}$W and a survey of one year.  For doing so, we have used the calculated values of $D_1$ for the two binary formation channels.  Given that the GW strain is inversely proportional to the distance of the source, the EM power released in the detector is also inversely proportional to the distance, see Eq.~(\ref{estim}).  Our final results are displayed in Fig.~\ref{fig_fDM}.    We can clearly see that the resonance plays a role by boosting the detection limits to be as low as $\tilde f_{\rm PBH} \lesssim 10^{-8}$ for primordial binaries, and $\tilde f_{\rm PBH} \lesssim 10^{-4}$ for tidal capture in clusters. .  Resonant EM-based HFGW detectors could therefore set unprecedented and independent limits on the abundance of planetary-mass PBHs.


 %
 
  \section{Conclusion} \label{sec:conc}

The detection of dozens of black hole or neutron star mergers by LIGO/Virgo and, more recently, the plausible observation of a stochastic GW background with pulsar timing arrays (PTAs) by NANOGrav, have revealed the bright future of GW astronomy.
LIGO/Virgo probe GW frequencies between $10$ and $10^4$ Hz and PTAs of $10^{-9}$ Hz.   In the future, ground-based and space-detectors like Einstein Telescope, Cosmic Explorer and LISA will fill the gap between these two ranges and probe other GW sources like intermediate mass black hole binaries.   On the other side of the GW spectrum, HFGW detectors might be equally interesting and probe cosmological stochastic backgrounds or exotic planetary-mass compact objects like PBHs.  Motivated by recent claims that PBHs with a wide mass distribution could explain, among others, the masses and spins of LIGO/Virgo black holes, the NANOGrav observation and the dark matter in the Universe, we envision their detectability with two designs of resonant EM detectors based on the inverse Gertsenshtein effect and operating at frequencies of order $100$ MHz.

The released energy in a EM resonator has been calculated for fourth-order post-Newtonian GW waveforms corresponding to the final inspiraling phase of light PBH binaries.   The experimental apparatus either consists in a one-meter long cylindrical resonant cavity (TM) or in a conducting waveguide (TEM), together with a static 5 Tesla magnetic field orthogonal to their axis of symmetry.    For a $5$ meters radius resonator or a waveguide of $0.1$ meter inner radius, we find a typical power variation of order of $10^{-10}$ W, detectable with present technology.  Similar resonant cavities have been already built for axion searches, but their collinear magnetic field prevents the detection of HFGWs. We have also studied the expected signal and its power spectrum for different binary masses. Their merging rates have been estimated for two PBH binary formation channels, at formation in the early Universe or due to tidal capture in dense clusters.  The optimal sensitivity is obtained for PBH masses of order $10^{-5} M_\odot$.  For the two considered models, we forecast the stringent limits on the PBH abundance that could be set with such experiments, of order $\tilde f_{\rm PBH} \lesssim 10^{-8}$ and $\tilde f_{\rm PBH} \lesssim 10^{-4}$, respectively.   Finally, one should keep in mind that we can tune the parameters of the detector to match the frequency range of highest interest.

Therefore HFGW therefore have the ability to set new constraints on the fraction of dark matter made of light PBHs that would be complementary to microlensing limits.  In particular, they could be used to distinguish between PBH and planetary origins of recently detected microlensing events towards the galactic center, suggesting that about one percent of the DM could be made of such objects. 

HFGW detectors with similar designs could also be used to detect cosmological stochastic GW backgrounds.   Frequencies around $100$ MHz are of particular interest in this context, because these are characteristic to GW sources at energies close to the GUT scale, like (p)reheating, oscillons, phase transitions or evaporating PBHs.   In particular, we point out that a strain sensitivity down to $h \sim 10^{-30}$ for achievable TM and TEM detectors is of order of the typical amplitude of such cosmological backgrounds.   Ongoing research focuses on using a frequency analysis of the resonance mechanim to get direct information about the frequency sensitivity of our detector. Another ongoing work is computing the power variation in the detectors generated by a stochastic, almost isotropic GW background.   
Other future research topic is to characterize the expected experimental noise with such detectors. The results for axion haloscope detection \cite{ADMX,ADMX2} could be an interesting starting point.  

In summary, resonant electromagnetic HFGW detectors are ideal to probe various aspects of fundamental physics, from early Universe cosmology to exotic compact objects like planetary-mass primordial black holes.   Our work contributes to pave the way in this direction and provides strong motivations to start the experimental development of such detectors, with currently available technology.\\\\

\noindent \textit{Acknowledgments}\\
This research used resources of the "Plateforme Technologique de Calcul Intensif (PTCI)"
(\url{http://www.ptci.unamur.be}) located at the University of Namur, Belgium, which is supported
by the F.R.S.-FNRS under the convention No. 2.5020.11. The PTCI is member of the "Consortium des Équipements de Calcul Intensif
(CÉCI)" (\url{http://www.ceci-hpc.be}).

\appendix
\renewcommand{\theequation}{\Roman{equation}}

\setcounter{equation}{0}
\section*{Appendix : Details of computations from Eq.(\ref{lineq}) to Eqs.(\ref{osc}-\ref{shat})}
In the Lorenz gauge, an incoming gravitational plane wave can be written down, in cylindrical coordinates $(t,r,\phi,z)$
%
\begin{eqnarray*}
h_{rr} (\phi,z,t)&=&-h_{\phi\phi}=\hp(z,t) \cos(2\phi) + \hx(z,t) \sin(2\phi)\\
h_{r\phi} (\phi,z,t)&=&-\hp(z,t) \sin(2\phi) + \hx(z,t) \cos(2\phi)
\end{eqnarray*}
with $h_{rr},h_{\phi\phi},h_{r\phi}$ are the metric perturbations in the non-coordinate basis
$(cdt,dr,r\,d\phi,dz)$ and 
where $\hp$ and $\hx$ are the usual polarizations of the incoming GW in the traceless-transverse gauge.

The incoming GW will interact with the component of the magnetic field that is perpendicular to its direction of propagation. We therefore consider the outer magnetic field along the x-direction : $\vec{B^{(0)}_{\rm ext}}=B^{(0)}_{\rm ext}\vec{e_x}\cdot$ Any component of the magnetic field that would be collinear to the direction of the propagation of the gravitational plane wave, $\vec{B^{(0)}_{\rm ext}}\approx \vec{e_z}$, does not produce any EM wave through the inverse Gertsenshtein effect. Indeed,  since 
$\tensud{F}{(0)}{xy}=B_z$, one has that, from Eq.(\ref{Smunu})

$$
S_{\mu\nu}=- \pr_x \left(\pr_\mu \tensd{h}{y \nu}-\pr_\nu \tensd{h}{y \mu}\right) \tensu{F}{(0)  \, x y}+\left(x\leftrightarrow y\right)\cdot
$$
Each term in the above equation will identically vanish in the case of a plane gravitational wave propagating along the direction of the magnetic field ($h_{+,\times}= h_{+,\times}(z,t)$).

Let now assume our EM resonators are axially symmetric along the $z-$direction and examine Eq. (\ref{lineq}) in cylindrical coordinates. The external constant magnetic field along the $x-$direction gives rise to only two components of the
zeroth order electromagnetic tensor 
\begin{eqnarray*}
\tensud{F}{(0)}{r z}&=B^{(0)}_{\rm ext} \sin(\phi)\\
\tensud{F}{(0)}{\phi z}&=B^{(0)}_{\rm ext} \cos(\phi)
\end{eqnarray*} 
in the non-coordinate basis $(cdt,dr,r \, d\phi,dz)\cdot$
For an incoming gravitational plane wave interacting with a constant magnetic field along the x-direction,
Eq.(\ref{lineq}) can be written down in terms of the induced electric and magnetic fields $\vec{E}^{(1)},\vec{B}^{(1)}$
\begin{eqnarray}
\left(-\frac{\partial^2}{\pr t^2}+\vec{\Delta}\right) \vec{E}^{(1)}=\vec{S}_E\label{waveE}\\
\left(-\frac{\partial^2}{\pr t^2}+\vec{\Delta}\right) \vec{B}^{(1)}=\vec{S}_B,\label{waveBA}
\end{eqnarray}
with the following source terms
\be
\vec{S}_E^\perp=B^{(0)}_{\rm ext} \frac{\partial^2  \hp}{\partial t \partial z } 
\begin{pmatrix}
-\sin(\phi)\\
-\cos(\phi)
\end{pmatrix}
+B^{(0)}_{\rm ext} \frac{\partial^2  \hx}{\partial t \partial z }
\begin{pmatrix}
	\cos(\phi)\\
	-\sin(\phi)
\end{pmatrix},
\label{E1par}
\ee
\be
\vec{S}_B^\perp=
B^{(0)}_{\rm ext}\frac{\partial^2  \hp}{\partial z^2 }
\begin{pmatrix}
	\cos(\phi)\\
	-\sin(\phi)
\end{pmatrix}
+B^{(0)}_{\rm ext}\frac{\partial^2  \hx}{\partial z^2 }
\begin{pmatrix}
	\sin(\phi)\\
	\cos(\phi)
\end{pmatrix}
\cdot\label{B1perp}
\ee
These Eqs.(\ref{waveBA}) and (\ref{B1perp}) leads to the Eq.(\ref{waveB}) of the paper.

Our demonstration earlier is illustrated here, indeed the longitudinal component $(E,B)^{(1)}_z$ are not sourced since $S_{E,z}=S_{B,z}=0$ so that the induced
radiation is purely transverse electromagnetic (TEM): $(E,B)^{(1)}_z=0$.
In the above equations, $\vec{\Delta}$ is the vector laplacian operator in cylindrical coordinates, which is given by
\be
\vec{\Delta}\vec{E}=
\begin{pmatrix}
\displaystyle{\nabla^2 E_r- \frac{E_r}{r^2}-\frac{2}{r^2}\pr_\phi E_\phi}\\
\\
\displaystyle{\nabla^2 E_\phi- \frac{E_\phi}{r^2}+\frac{2}{r^2}\pr_\phi E_r}\\
\\
\nabla^2 E_z
\end{pmatrix}
\ee
where $\nabla^2$ is the scalar laplacian : $\nabla^2 f=\pr_r^2 f+\frac{1}{r}\pr_r f+\frac{1}{r^2}\pr^2_\phi f+\pr_z^2 f\cdot$.
Any EM field inside an ideal resonator must verify the boundary conditions along any perfect conducting surfaces: $\vec{E}^{(1)}_\parallel=\vec{B}^{(1)}_\perp=\vec{0}\cdot$ Therefore, any waveguide with two concentric open cylinders will host EM field configurations with $E^{(1)}_{z}|_{r=R1,R_2}=B^{(1)}_{r}|_{r=R_1,R_2}=0$ while a cylindrical conducting cavity will host EM fields satisfying
$E^{(1)}_{z}|_{r=R}=B^{(1)}_{r}|_{r=R}=0$ and $E^{(1)}_{r,\phi}|_{z=\pm L/2}=B^{(1)}_{z}|_{z=\pm L/2}=0\cdot$.
The inhomogeneous wave equations Eqs.(\ref{waveE},\ref{waveBA}) can be efficiently solved using spectral methods, i.e. by approximating the unknown field, $(\vec{E},\vec{B})^{(1)}(t,r,\phi,z)$ by a decomposition over a basis of orthogonal functions that satisfy the above-mentioned boundary conditions. Here, a natural choice of such basis is given by the following set of cylindrical harmonics (omitting the useless longitudinal component $\psi^z_{kmn}$)
\be
\psi^r_{kmn}=C_{kmn}\cdot\mathcal{R}_{km}(r)\cdot 
\begin{Bmatrix}
\cos\\
\sin
\end{Bmatrix}
\left(m\phi\right)
\cdot 
\begin{Bmatrix}
\cos\\
\sin
\end{Bmatrix}
\left(\frac{2\pi n z}{L}\right)
\ee
\be
\psi^\phi_{kmn}=\mp C_{kmn}\cdot\mathcal{R}_{km}(r)\cdot 
\begin{Bmatrix}
\sin\\
\cos
\end{Bmatrix}
\left(m\phi\right)
\cdot 
\begin{Bmatrix}
\cos\\
\sin
\end{Bmatrix}
\left(\frac{2\pi n z}{L}\right)
\ee

where $C_{kmn}$ are normalization coefficients and where the radial eigenfunctions $\mathcal{R}_{km}(r)$ are such that the boundary conditions are satisfied. This is the case when
\bea
\mathcal{R}_{km}(r)&=&A_{k}.J_{m-1}(\alpha_{k}.r)+Y_{m-1}(\alpha_{k}.r)\\
&=&J_{m-1}(\alpha_{k}.r),
\eea
for the waveguide and the cavity, respectively.
In the above equation, $J_n(r),Y_n(r)$ are Bessel functions of the first and second kind, respectively, and the constants $A_{km}$ and $\alpha_{k}$ are solutions of the following system
\bea
A_{k}.J_{m-1}(\alpha_{k}.R_{1,2})+Y_{m-1}(\alpha_{k}.R_{1,2})&=&0,
\label{freq1}
\eea
for the case of the waveguide with two open conducting cylinders and
\bea
J_{m-1}(\alpha_{k}.R)&=&0,
\label{freq2}
\eea
in the case of the cylindrical cavity.
 
 Those cylindrical harmonics $\left(\vec{\psi}_{kmn}\right)^T=\left(\psi^r_{kmn},\psi^\phi_{kmn},\psi^z_{kmn}\right)$ are eigenfunctions of the vector laplacian operator 
 $ \vec{\Delta}$ in
 cylindrical coordinates with the following eigenvalues:
 \be
 \vec{\Delta} \vec{\psi}_{kmn}=-\left(\alpha^2_{k}+\frac{4\pi^2 n^2}{L^2}\right)\vec{\psi}_{kmn}=-\Omega^2_{kn}\vec{\psi}_{kmn}
 \ee
These eigenfunctions $\psi^{r,\phi}_{kmn}$ constitute a complete orthonormal set with scalar product
 $$
 \left(f,g\right)=\int_{R_1(0)}^{R_2(R)}\int_0^{2\pi} \int_{-L/2}^{L/2} f(r,\phi,z)g(r,\phi,z) r.dr.d\phi. dz
 $$
 where the different bounds of the integral over the radius correspond to the case of a waveguide or a cavity.

 Moving back to Eqs.(\ref{waveE},\ref{waveBA}),
 one can use a truncated expansion on this basis as an approximation of the unknown field. For instance, let us set 
 \be
 B^{(1)}_{r,\phi}(t,r,\phi,z)\approx \sum_{k,m,n} \hat{b}^{r,\phi}_{k,m,n}(t) \cdot\psi^{r,\phi}_{kmn}(r,\phi,z)
 \ee
 with
 $$
 \hat{b}^{r,\phi}_{k,m,n}(t)=\left(B^{(1)}_{r,\phi},\psi^{r,\phi}_{k,m,n}\right)
 $$
 so that the inhomogeneous wave equation Eq.(\ref{waveBA}) 
  now reduces to a harmonic oscillator ODE for each mode $\hat{b}^{r,\phi}_{k,m,n}(t)$, Eq.(\ref{osc}) in the paper.
 \be
 \frac{d^2\hat{b}^{r,\phi}_{k,m,n}}{dt^2}+\Omega^2_{kn}\hat{b}^{r,\phi}_{k,m,n}=\hat{s}^{r,\phi}_{k,m,n}(t)
 \tag{\ref{osc}}
 \label{oscA}
 \ee
 where $\hat{s}^{r,\phi}_{k,m,n}(t)$ are the spectral coefficients of the source $S^{r,\phi}_B$:
 $$
S_{B}^{r,\phi}(t,r,\phi,z)\approx \sum_{k,m,n} \hat{s}^{r,\phi}_{k,m,n}(t) \cdot\psi^{r,\phi}_{kmn}(r,\phi,z)\cdot
 $$
In this approach, each mode $\hat{b}^{r,\phi}_{k,m,n}(t)$ that composes the induced magnetic field $\vec{B}^{(1)}$
behaves as an harmonic oscillator driven by the corresponding mode of the source $\hat{s}^{r,\phi}_{k,m,n}$. 
The general solution of such system describes a resonance process through a superposition of oscillations at proper frequency $\Omega_{kn}$ and a particular solution of the inhomogeneous equation. This last can easily be obtained with a Fourier series method. 
Since the source of the induced magnetic field is directly related to the second derivative of the metric perturbation, resonance will be achieved when the incoming GW share some common frequencies with the coaxial detector, as we have seen in the main text. 

Fortunately, there are some simplifications allowing to restrict to specific modes of the cylindrical harmonic decomposition of the induced EM fields $\left(\vec{E}^{(1)},\vec{B}^{(1)}\right)\cdot$ First, in the case of a plane gravitational wave with a direction of propagation perpendicular to the one of the outer magnetic field $\vec{B}^{(0)}$, the source terms are given by Eqs.(\ref{E1par}-\ref{B1perp}) and therefore the induced EM fields will only have an azimuthal mode number equal to one: $m=1$. Second, if ones focuses on the detection of the induced EM energy inside the resonator then it suffices to consider the components of the induced magnetic fields that are in the same direction than the outer strong magnetic field. Indeed, the EM radiation that is induced into the resonators will modify their EM energy content. The total electromagnetic energy $\mathcal{E}$ inside the waveguide is given by the classical formula
$$
\mathcal{E}=\frac{1}{2}\int_V \left(\epsilon_0 ||\vec{E}||^2+\frac{||\vec{B}||^2}{\mu_0}\right)dV
$$ 
with $V$ the volume of the waveguide. The dominant contribution to the energy variation $\Delta \mathcal{E}$
is given by the coupling between the external magnetic field $\vec{B}^{(0)}$ and the induced  one
$\vec{B}^{(1)}$
\be
\Delta \mathcal{E}=E_{\rm tot}-E^{(0)}\approx \frac{1}{\mu_0}\int_V \left(\vec{B}^{(0)}\bullet\vec{B}^{(1)}\right)dV
\ee
neglecting the terms in $||\vec{E}^{(1)}||^2$ and in $||\vec{B}^{(1)}||^2\cdot$ 
Since we have assumed here the external magnetic field lies in the 
$x-$direction, one only needs to compute
$$
B^{(1)}_{x}=B^{(1)}_{r}\cos\phi-B^{(1)}_{\phi} \sin\phi
$$
from Eqs.(\ref{waveBA},\ref{B1perp}). In the case of incoming plane GW orthogonal to the outer magnetic field, only the modes with $m=1$ (such that $B^{(1)}_{r}\approx \cos\phi$ and $B^{(1)}_{\phi}\approx -\sin\phi$, see above)  and  $n=0$, i.e. constant in the $z-$direction, will give a non-vanishing contribution to the integral over the volume in the computation of the energy. The important terms are therefore sourced by the Eq.(\ref{shat}), 
\be
\hat{s}_{k,1,0}^{r,\phi}(z,t)=\pi B_0 L^2 \mathcal{I}_k\int_{-L/2}^{L/2} \frac{\partial^2  \hp(z,t)}{\partial z^2 }dz
\tag{\ref{shat}}
\label{shatA}
\ee
with the dimensionless quantity $\mathcal{I}_k$ given by
\be
 \mathcal{I}_k=\int_{r_1(0)}^{r_2(R/L)}\mathcal{R}_{k,1}(L\rho) \rho  d\rho
\ee
with $\rho=r/L$, $r_{1,2}=R_{1,2}/L$  and
for a magnetic field along the $x-$direction.

 In summary, the energy variation is given by, at the leading order, the Eq. (\ref{deltaE}) 
\be
\Delta \mathcal{E}\approx \frac{2\pi B_0 }{\mu_0} \cdot \sum_{k} \mathcal{I}_k\hat{b}_{k,1,0}(t)
\tag{\ref{deltaE}}
\label{deltaEA}
\ee
with $\hat{b}_{k,1,0}(t)=\hat{b}_{k,1,0}^r(t)+\hat{b}_{k,1,0}^\phi(t)=2\hat{b}^{r,\phi}_{k,1,0}(t)\cdot$
Those modes $\hat{b}_{k,1,0}(t)$ are solutions of the forced harmonic oscillator equation (\ref{oscA})
 with source term given by (twice) Eq.(\ref{shatA}). This completes the detailed explanation to obtain Eqs.(\ref{osc}-\ref{shat}).

\bibliographystyle{apsrev4-2}
\bibliography{biblio} 

\end{document}